\pdfoutput=1

\documentclass[12pt, preprint]{aastex}

\shorttitle{Hot Plasma in Non-Flaring Active Regions}
\shortauthors{Ko et al.}

\newcommand{\lsim}{\lower.5ex\hbox{$\; \buildrel < \over \sim \;$}}
\newcommand{\gsim}{\lower.5ex\hbox{$\; \buildrel > \over \sim \;$}}

\begin{document}

\title{Hot Plasma in Non-Flaring Active Regions Observed by the 
Extreme-ultraviolet Imaging Spectrometer on {\it Hinode}}

\author{Yuan-Kuen Ko\altaffilmark{1}, George A. Doschek\altaffilmark{1}, Harry P. Warren\altaffilmark{1}, Peter R. Young\altaffilmark{1,2}}
\altaffiltext{1}{Space Science Division, Naval Research Laboratory, Washington, DC 20375, USA}
\altaffiltext{2}{George Mason University, 4400 University Drive, Fairfax, VA 22030, USA} 

\email{yko@ssd5.nrl.navy.mil}

\begin{abstract}

The Extreme-ultraviolet Imaging Spectrometer (EIS) on the {\it Hinode}
spacecraft obtains high resolution spectra of the solar atmosphere in
two wavelength ranges: 170 - 210 and 250 - 290 \AA.  These wavelength
regions contain a wealth of emission lines covering temperature
regions from the chromosphere/transition region (e.g., \ion{He}{2},
\ion{Si}{7}) up to flare temperatures (\ion{Fe}{23},
\ion{Fe}{24}).  Of particular interest
for understanding coronal heating is a line of \ion{Ca}{17} at 192.858 \AA, formed
near a temperature of $6\times10^6$ K.  However, this line is blended with
two \ion{Fe}{11} and six \ion{O}{5} lines.  
In this paper we discuss a specific procedure to extract the Ca XVII line from the blend. 
We have performed this procedure on the raster data of five active regions and a limb flare, and demonstrated that the Ca XVII line can be satisfactorily extracted from the blend if the Ca XVII flux contributes to at least $\sim$ 10\% of the blend. We show examples of 
the high-temperature
corona depicted by the Ca XVII  emission and find that the Ca XVII emission has three morphological features in these active regions -- 1) `fat' medium-sized loops confined in a smaller space than the 1 million degree corona, 2) weaker, diffuse emission surrounding these loops that spread over the core of the active region, and 3) the locations of the strong Ca XVII loops are often weak in line emission formed from the 1 million degree plasma. 
We find that the emission measure ratio of the 6 million degree plasma relative to the cooler 1 million degree plasma in the core of the active regions, using the Ca XVII to Fe XI line intensity ratio as a proxy, can be as high as 10. Outside of the active region core where the 1 million degree loops are abundant, the ratio has an upper limit of about 0.5. 

\end{abstract}


\section{Introduction}

The Extreme-ultraviolet Imaging Spectrometer (EIS, Culhane et al. 2007) on the {\it Hinode}
spacecraft (Kosugi et al. 2007) measures coronal spectral line intensities, profiles,
and Doppler shifts in two wavelength ranges: 170 - 210 and 250 - 290 \AA. 
These wavelength regions contain a wealth of emission lines covering temperature
regions from the chromosphere/transition region (e.g., \ion{He}{2},
\ion{Si}{7}) up to flare temperatures (\ion{Fe}{23},
\ion{Fe}{24}) (Young et al. 2007, Brown et al. 2008). It also has the capability of imaging solar
features by rastering the spectrometer slit over the region, forming
images by integrating over the line profiles. For resolved lines (i.e. non-blend, or blended but extractable
from line fitting procedures), these images generally represent emission from around their
respective formation temperatures (e.g. Doschek et al. 2007a).
Thus EIS has better temperature discrimination than broadband imaging
telescopes that form images from spectral responses that contain lines
formed over a range of temperatures.  Since launch EIS has obtained
such high spatial resolution spectral images for many active regions
in spite of being launched near solar minimum.

The EIS spectra give us the opportunity of obtaining accurate
differential emission measures (DEM) at different locations in the active
region. This is important for solar irradiance applications (e.g. Warren et al. 2001) and coronal heating models (e.g. Klimchuk 2006).  
The construction of DEMs are bounded by the lowest and highest
temperature lines that EIS can record.  
EIS can observe an abundance of lines at
temperatures as low as a few hundred thousand degrees and as high as
about $20\times10^6$ K (e.g. Young et al. 2007, Brown et al. 2008).  The highest temperature lines 
available, from \ion{Fe}{24} and \ion{Fe}{23}, are only seen during flares.
There are several relatively high temperature lines, with formation temperature
higher than 3 million degrees, that can appear in active regions 
in the absence of obvious flaring.  
As recently analyzed by Warren et al. (2008b), these lines are mostly
from Fe XVII and Ca XIV,XV,XVI,XVII. Among them, Fe XVII and Ca XVII lines are
of the highest formation temperature at $5-6 \times10^6$ K, and are
of particular interest in that they can provide much needed constraints
in coronal heating models (Parenti et al. 2006; Patsourakos \& Klimchuk 2007).

The Ne-like \ion{Fe}{17} lines are
excited from the ground state to the first excited levels, which
involve energies in the X-ray region equivalent to wavelengths around
15-17 \AA.  Many of the excited states have decay channels back to the
ground state and produce strong \ion{Fe}{17} X-ray lines near the
above wavelengths that are well-known features in the solar X-ray
spectrum as well as in the spectra of other stars and more exotic
astrophysical objects.  However, due to branching ratios and the fact
that two of the lines arise from states where a return to the ground
state is strictly forbidden by radiative decay, there are some
\ion{Fe}{17} lines that appear within the EIS wavebands.  These lines
were extensively studied during the {\it Skylab} era and spectroscopic
results are summarized by Feldman, Doschek, \& Seely (1985) and
Doschek, Feldman, \& Bhatia (1991). 
Unfortunately, the EUV \ion{Fe}{17} lines are quite weak except during
flares, and were not included in many of the EIS active region raster
scans made up to now. The detailed analysis of Warren et al. (2008b) concluded
that it is difficult at the current stage to use the Fe XVII spectra for quantitative analysis
due to problems in atomic physics.

The Be-like \ion{Ca}{17}
line is relatively strong and is one of the lines included in most
observations made with EIS. According to the CHIANTI atomic database version 5.2 
(Dere et al. 1997; Landi et al. 2006), the emissivity of Ca XVII $\lambda$192.858 is about a factor 
of 4 higher than that of the strongest Fe XVII line (204.65 \AA) observable by EIS at their peak formation
temperature ($5-6 \times10^6$ K, see also Figure 4 of Warren et al. 2008b). 
The observed Ca XVII can even be at least a factor 20 stronger than Fe XVII when also taking into account their instrument effective areas. 
However, it is blended with two \ion{Fe}{11} and six \ion{O}{5} lines. For certain data such as in flare loops, the Ca XVII line
dominates the emission of the blend (e.g. Hara et al. 2008).  In most situation such as in non-flaring active regions,
the contribution from Fe XI (and O V at certain locations) usually dominates, or is comparable to, the Ca XVII emission.
Previous studies usually obtained the Ca XVII line intensity by subtracting some estimated Fe XI and O V contribution
from the whole blend (e.g. as suggested by Young et al. 2007). 
This method works well in general but can produce large errors at where Fe XI, Ca XVII and O V emissions are
comparable with one another. 
In this paper we present the first attempt to obtain the Ca XVII line by multi-Gaussian 
fitting of the entire profile of the blend. We believe that this specific procedure provides the best way to date for
obtaining this Ca XVII line under a wide variety of conditions. This \ion{Ca}{17} line, combining with other Ca lines at lower ionization stages as suggested by Warren et al. (2008b),  is therefore the most practical high temperature line available to probe the high temperature corona in the absence of flares.

In Section 2 we briefly describe the EIS spectrometer.  In Section 3
we discuss the blending of the \ion{Ca}{17} complex and the method to remove
the Fe XI and O V components from the Ca XVII component.  We show results of
this procedure for five active regions and a limb flare in Section
4. In Section 5 we discuss the morphology and emission measure of the 6 million degree corona 
relative to the cooler plasma in these active regions. We give our summary in Section 6.

\section{The EIS on {\it Hinode}}

The EIS is described in detail by Culhane et al. (2007) and Korendyke
et al. (2006).  The instrument consists of a combination multi-layer
telescope and spectrometer.  The telescope mirror and spectrometer
grating are divided into two segments, each of which is coated with
different Mo/Si multi-layers tuned to different wavelength bands:
170-210 \AA~and 250-290 \AA.  Light focused from the telescope onto
the entrance aperture of the spectrometer enters the spectrometer and
is then diffracted by the grating and focused onto two CCD detectors.

The telescope mirror is articulated and different regions of the Sun
can be focused onto the spectrometer aperture by fine and coarse
mirror motions.  The entrance aperture of the spectrometer has four
options: a 1\arcsec\ slit, a 2\arcsec\ slit, a 40\arcsec\ slot, or a
266\arcsec\ slot.  The slits/slots are aligned in the solar
north/south direction.  The CCD height is 1024\arcsec. The heights for individual observations
can be varied, with a maximum height for most observations to be 512\arcsec.

EIS can be operated in several modes.  Images of solar regions can be
constructed by rastering a slit or slot in the west to east direction
across a given solar area with a set exposure time at each step.  At
each raster position it is possible to obtain a complete spectrum for
each wavelength band.  However, it is also possible to select a small
set of lines with specified spectral windows, the choices
of lines depending on the objectives of the observation.
The spatial resolution of EIS along the slit and in the direction of
dispersion is approximately 2\arcsec\ (1\arcsec\ per pixel).  The
spectral dispersion is 0.0223 \AA\ per pixel.  
The instrumental full width at half maximum (FWHM) of a line profile measured in
the laboratory prior to launch is 1.956 pixels.  
Dosheck et al (2007b, 2008) adopted a FWHM width  
of 0.056 \AA, similar to that found by Brown et al. (2008). 

Prior to detailed data analysis, the level-0 data were calibrated by the standard calibration routines that remove the pedestal and dark current, electron spikes and hot pixels. Wavelengths were corrected for the slit tilt and thermal variations during the orbit. We refer Young et al. (2009) for details. 

\section{Data Reduction}

The main purpose of this work is to extract the \ion{Ca}{17} 192.858 \AA\ line from the blending with the \ion{O}{5} and \ion{Fe}{11} lines. 
Table 1 lists the six \ion{O}{5} transitions, two \ion{Fe}{11} transitions and the \ion{Ca}{17} transition in the $\lambda$192 blend that span a wavelength range of 0.165 \AA. Also listed are the two Fe XI lines in the $\lambda$188 blend. To demonstrate the blend, we use the observation of AR10978 on December 11, 2007. The upper panels of Figure 1 show the raster images of the $\lambda$192 blend (upper left) and the Fe XI $\lambda$188.216 line (upper right). We can see that the two images are very similar due to the predominant Fe XI emission almost everywhere in the $\lambda$192 blend. However, the contrast among some structures is different in the two images, indicating different contributions from the O V and Ca XVII lines relative to the Fe XI emission. For example, the locations marked by `1' and `2' do not stand out in the Fe XI $\lambda$188.216 image as in the image of the $\lambda$192 blend, indicating significant emission from O V or Ca XVII, and structures around point `3' are especially bright in the Fe XI image indicating that the Fe XI emission dominates. The six plots below the two raster images are the spectra at the three marked locations for the $\lambda$192 blend (middle panels) and the Fe XI $\lambda$188 blend (lower panels). Indeed, Point `1' shows a significant O V emission (a second peak is at the right of the Fe XI line). Point `2' shows a likely significant Ca XVII emission judged from the broad width of the line profile due to the combination of Fe XI and Ca XVII, as well as that the line peak is close to the rest wavelength of Ca XVII but there is no comparable line shift in the Fe XI $\lambda$188.216 line. Point `3' shows that the Fe XI emission is dominant. Note that the line spectra seem to lie above different background levels at the three locations. This is mainly due to different intensities of the neighboring lines (e.g. Fe XI 192.62 \AA\ and Fe XIV 192.63 \AA\ are at the short wavelength side. See Brown et al. 2008). Most of the instrument background is already subtracted by the standard calibration procedure (see Young et al. 2009 for details). 

\subsection{Procedure for extracting the Ca XVII line from the blend}

With 9 lines in the blend and if we assume each line can be represented by a Gaussian, there would be 28 parameters to fit with (peak, width and centroid for each line, plus a constant background level). However, the number of free parameters can be reduced considerably. With simultaneous observation of the Fe XI $\lambda$188.216/$\lambda$188.299 lines which are bright and easy to fit with a two-Gaussian profile (see lower plots of Fig.1), the spectral profiles of the two Fe XI $\lambda$192 lines are essentially determined given that the rest wavelengths and their relative line ratios are known (see below), and that the line width and Doppler shift of all Fe XI lines are the same. 
Similarly, the combined profile of the six O V lines (i.e. six Gaussians) can be calculated from just 3 parameters: peak, width and centroid of the strongest O V 192.904 \AA\ by prescribed line separations and ratios (see below). Adding in the Gaussian profile of the Ca XVII line (3 parameters) and a constant background level, the number of free parameters for fitting the entire blend is thus reduced to only seven. A least-squares minimization algorithm (MPFIT) is used to fit the data and obtain the best-fit parameter values of the 7 free parameters. Note that for the data we analyzed here, we chose the background being at a constant level. This would contribute to some systematic error in the fitted results since the blending by line wings of adjacent lines may give a false background level at either side of the concerned lines (e.g. see Fig.1). Young et al. (2009) adopted a method of identifying the background level but it requires a large spectral window. For most data we analyzed here, the concerned spectral windows are too narrow for us to do so. Nonetheless, we believe that the error caused by this assumption should be a small contribution to the overall uncertainty of the fitting procedure. 

Below we first describe in detail the methods and reasonings that arrive at our procedure of extracting the Ca XVII line out of the blend. A summary is then provided listing out individual steps of the procedure. 

The rest wavelengths (see Table 1) of the six O V lines and line intensity ratios relative to O V 192.904 \AA\ are taken from CHIANTI atomic database version 5.2 (Dere et al. 1997, Landi et al. 2006), specifically, from Fuhr et al. (1999), Tachiev \& Froese Fischer (1999), and K. A. Berrington (2003, private communications). These O V line ratios are not sensitive to the electron density below 10$^{10}$ cm$^{-3}$ and the ratios increase by 25\% for the 192.75 \AA, 192.801 \AA\ and 192.915 \AA\ lines, and by 50\% for the 192.797 \AA\ and 192.911 \AA\ lines at $n_e$=10$^{12}$ cm$^{-3}$. For this work, we take the line ratios to be the average between $n_e$ of 10$^{8}$ and 10$^{10}$ cm$^{-3}$ (at the O V peak formation temperature of  $2.5\times10^5$ K). They are 0.1729, 0.3193, 0.1296, 0.1063, 0.00863 for the 192.75 \AA, 192.797 \AA,192.801 \AA, 192.911 \AA, and 192.915 \AA\ lines, respectively. The standard deviations are only 1-2 \% of the mean value. 

The rest wavelengths (Table 1) of the Fe XI lines are taken from Brown et al. (2008). These adopted rest wavelengths are found to be consistent with the EIS data when compared with the line centroids from the fitting. 
The intensity ratios of the Fe XI lines, however, need special attention. The difficulty in calculating accurately the atomic data for Fe XI is known and has been extensively studied by Young (1998).  According to Young (1998), Fe XI $\lambda$188.299 is one of the most troublesome transitions due to complicated coupling of its upper level (see Table 1) with other levels. This would be the same situation for Fe XI $\lambda$192.901 which has the same upper level. Therefore we choose to determine these line ratios from the EIS data. To do so, we studied three EIS off-limb quiet Sun observations on Mar.9, 2007, Mar.11, 2007, and Sep.26, 2007.  For each data set, we selected a spatial region off the solar limb that is free of obvious extended loops, then fitted the data with two Gaussians for both the $\lambda$188 and $\lambda$192 blends. For the $\lambda$192 blend, it is reasonable to assume that the signal off limb should come dominantly from the two Fe XI lines because O V lines originate from the transition region lower down and high temperature {\it non-flaring} loops are generally believed to be low-lying and located only in the core of active regions (e.g. see Warren et al. 2008b). The fitting is performed for each pixel in the selected region. We find that the most probable ratio is 0.70 and 0.26 for 188.299/188.216 and 192.813/188.216, respectively. The scatter of the resulting 192.901/192.813 ratios is large. We take the average of the three median ratios for the 3 datasets and adopt the 192.901/192.813 ratio of 0.02. These empirical values can be compared with those in the CHIANTI database v5.2 of 0.36, 0.21, and 0.084 for 188.299/188.216 and 192.813/188.216, and 192.901/192.813, respectively (at $1.26\times 10^6$ K, and $10^9$ cm$^{-3}$). The large discrepancy in the 188.299/188.216 and 192.901/192.813 ratios is obvious.  Current efforts are already underway to refine the calculation for this ion (see Young et al. 2007). Until more accurate atomic calculations are available for this ion, our choice of these intensity ratios is essentially empirical. One final note about the troublesome Fe XI atomic data is that CHIANTI v5.2 lists a theoretical line at 192.832 \AA\ (decay of level 3s$^2$3p$^3$3d $^3$S$_1$ to 3s$^2$3p$^4$ $^3$P$_1$) whose strength is predicted to be about 25\% (at $n_e=10^9$ cm$^{-3}$) that of Fe XI 192.813 \AA. We find that this wavelength is likely wrong because it would otherwise predict a line a factor of 2.4 stronger at 203.331 \AA\ (decay of level 3s$^2$3p$^3$3d $^3$S$_1$ to 3s$^2$3p$^4$ $^1$D$_2$) which should be observable by EIS. We checked some EIS data for quiet Sun, active region and off-limb observations and looked for this line. We find that this 203.331 \AA\ line either does not appear in the data, or has intensity less than 1/100 of the 188.216 \AA line. Therefore we conclude that the documented wavelength of 192.832 \AA\ for this theoretical line is not correct and this line should not be included in the $\lambda$192 blend.


We summarize the procedure into sequential steps as follows. 1)  Obtain the peak, width and centroid of Fe XI 188.216 \AA\ from 2-Gaussian fitting of the Fe XI $\lambda$188 blend.  2) The peaks of the Gaussians for Fe XI 192.813 \AA\ and Fe XI 192.901 \AA\ are then 0.26 and 0.0052 times that of Fe XI 188.216 \AA, respectively. The centroids have the same Doppler shift and the widths are the same as that of Fe XI 188.216 \AA. The combined Gaussian functions for the two Fe XI  $\lambda$192 lines are then calculated. 3) For a given Gaussian peak, width and centroid of the O V 192.904 \AA\ line, the Gaussian peaks of the 192.75 \AA, 192.797 \AA, 192.801 \AA, 192.911 \AA, and 192.915 \AA\ lines are 0.1729, 0.3193, 0.1296, 0.1063, 0.00863 that of the O V 192.904 \AA, respectively. All O V line widths and Doppler shifts are the same. The combined Gaussian functions for the six O V lines are then calculated. 4) Adding the Gaussian function for Ca XVII (specified by peak, width and centroid) and a constant background level, the best-fit values for the seven free parameters, thus the Ca XVII and O V line parameters, can be obtained by fitting the entire function with the data by a least-squares minimization algorithm (e.g. MPFIT). Note that the  adopted value of 0.26 for the Fe XI $\lambda$192.813 to $\lambda$188.216 ratio is consistent with that suggested in Young et al. (2007), and should be independent of the electron density because both transitions originate from the same upper level. This eliminates the uncertainty caused by variations in density across the active region.

In this work, the procedure is performed for each x-y pixel (1\arcsec\ by 1\arcsec\ area) of the raster data. In order not to obtain unrealistically large or small line widths from the fitting, the line widths (here we define as the 1/e half width of the Gaussian profile divided by $\sqrt 2$) are limited to the range of 0.02 \AA\ and 0.045 \AA\ (i.e. FWHM of 0.047 \AA\ and 0.106 \AA, respectively). 
The instrument width is around 0.024 \AA\ (i.e. FWHM of 0.056 \AA, see previous section). Thus 0.02 \AA\ is a reasonable lower limit. The upper limit of 0.045 \AA\ is also reasonable judging from our experience of fitting several Fe XI and Fe XVI lines for several active regions that the widths are rarely larger than 0.045 \AA. Because the rest wavelengths of Ca XVII and O V 192.904 \AA\ are only 0.04 \AA\ apart, in order not to confuse one line for the other during the fitting process, we limit the range of the centroid shift of both lines by $\pm$ 0.04 \AA\ ($\sim$60 km/s). Therefore special measures need to be taken if a given data point is known to have a very large Doppler shift in either line (e.g. judged by other lines at similar formation temperature). Note that we have tested this procedure under various assumptions, e.g. relieving the limits for line width or line centroid. We obtain similar results. The differences among them are especially small at locations where Ca XVII or O V lines are strong. This indicates the robustness of this procedure to obtain strong emission features from both lines.

\subsection{Uncertainty investigation for Ca XVII}

To understand the limitations of this deconvolution algorithm of the multi-Gaussian fitting, we have performed a series of
Monte Carlo simulations. In these simulations we have held the total counts in
the \ion{Fe}{11} 192.813\,\AA\ line fixed and varied the relative contribution of
\ion{Ca}{17} and \ion{O}{5} to the blend. For every profile we add Poisson noise to each
spectral pixel and pass the composite profile to the fitting routine exactly as it was
applied to the actual data. As a measure of the error we compute the ratio of computed to
input line intensity. An example Monte Carlo simulation is shown in Figure 2.
This process was repeated 1000 times for each value of relative \ion{Ca}{17}
and \ion{O}{5} intensity so that statistics on the errors can be accumulated.
The total \ion{O}{5} intensity is varied from 0.1 to 10 times the \ion{Fe}{11}
intensity while the \ion{Ca}{17} intensity is varied from 0.1 to 100 times the \ion{Fe}{11}
intensity. A total of 96,000 simulated profiles were fit.

The results of these simulations are shown in Figure 3. It suggests that when
the \ion{Ca}{17} intensity is comparable to the flux in the \ion{Fe}{11}
192.813\,\AA\ line the error is approximately 20\% or less. These simulations, however,
also indicate that it is not possible to extract the \ion{Ca}{17} intensity accurately
once it falls below about 10\% of the flux in the \ion{Fe}{11} 192.813 \AA\ line.  At
these levels the error in the \ion{Ca}{17} intensity is approximately 100\%. Note that the
error in the \ion{Ca}{17} intensity is less dependent of the \ion{O}{5}
contribution. We thus find that whether or not this \ion{Ca}{17} line can be successfully extracted from the blend
does not depend on the counting statistics alone. It also depends on how bright it is relative to the Fe XI and O V emission.
Similar situation applies to the extraction of O V in the blend. In the case of O V (lower two panels of Fig.3), the errors not only  depend on the relative intensity of O V to Fe XI, but also depend on the Ca XVII to Fe XI ratio, especially when the Ca XVII contribution is larger than Fe XI. Based on this study, we choose a criterion that if the Ca XVII (or O V) intensity out of the fitting is larger than 10\% of the whole blend, it is regarded as trustable.   
It should be remembered that these Monte Carlo simulations only account for statistical
variations. Potential systematic errors, such as the assumed relative intensities of the
\ion{O}{5} components or the assumed \ion{Fe}{11} 192.813 to 188.216\,\AA\ ratio, are not
included. 

\section{Active Region Corona at Six Million Degrees}

\subsection{December 11, 2007, AR 10978}

Figures 4 shows again the raster images of the $\lambda$192 blend (upper left panel) and Fe XI 188.216 \AA\  (upper right panel, which is equivalent to Fe XI $\lambda$192.813) for the December 11, 2007 data. 
In the middle panels are the images of Ca XVII 192.858 \AA\ (left) and O V 192.904 \AA\ (right) derived from this work. 
Based on the uncertainty estimate discussed in Section 3.2, we do not take into account those pixels that have Ca XVII (O V) flux less than 10\% of the combined flux in the blend (this applies to all figures that follow). Those pixels below this intensity limit are set to the minimum value in the plotted images. 
Six locations are marked to show examples of the fitting results in Figure 5. We pick these six locations to show where the emission is strong in O V (points `1' and `4'), Ca XVII (points `2' and `5'), or Fe XI (points `3' and `6'). We can see that the quality of the fit is very satisfactory. We did spot checks at many other locations and find similarly good fitting results. Therefore we are confident that this procedure for extracting Ca XVII (and O V also) from the blend is accurate (although limited by the uncertainties discussed in Section 3). 

The lower left panel shows a Hinode/XRT image that was taken at a time (17:13:20 UT) during the EIS raster observations, and the selected time is close to that at a raster X-position (X=-85") where the strong Ca XVII emission was seen. Note that this active region has been quiescent, producing no X-ray flares above B-class since it appeared from the east limb a week earlier. Nonetheless, the XRT images taken during this raster observation still exhibit notable changes in brightness among various locations but the overall structure is more or less the same. One also needs to keep in mind that a raster image is constructed by scanning the slit positions, therefore is not a snapshot at a given time. The lower right panel shows the raster image of Fe XVI $\lambda$262.976 (peak formation temperature at 2.5 million degrees, from the 1-Gaussian fits) for comparison. To illustrate the different morphology exhibited by different emission lines and XRT filters, thus hinting on possible differences in the temperature structure, Figure 6 plots the emissivities of Fe XI $\lambda$188.216, Fe XVI $\lambda$262.976, Ca XVII $\lambda$192.858, and the temperature response functions for the Be\_med and Al\_poly filters of Hinode/XRT. The curves are all normalized to their respective maximum.   

Obvious structures of the Ca XVII emission, appearing as large-scale loops, stand out which are otherwise hidden in the $\lambda$192 blend. On the other hand, the O V emission exhibits typical transition region/chromospheric network structure with several spotty small-scale brightenings within the active region. The Ca XVII loops appear more like `fat bundles' instead of thin threads like those seen in Fe XI. The length of these loops is comparable to the size of the active region `core' (i.e. area of sunspots where the EUV/X-ray emission is also the brightest), and is not as extended as some Fe XI loops that are commonly seen from the 1-2 million degree corona. This indicates that these high-temperature loops are low-lying and are associated with areas of strongest magnetic field strengths. There is some diffuse emission surrounding these Ca XVII loops. Outside of the loops and the diffuse regions, the Ca XVII emission appears featureless. Those featureless areas generally have Ca XVII fluxes smaller than 10\% of the blend. 

One interesting feature is that the locations with bright Ca XVII emission, e.g. at points `2' and `5', appear dark in the Fe XI emission. The two thin loops seen in Fe XI  just south of point '5' (upper right panel of Fig.4) appear to enclose the bright Ca XVII emission in between them. The bright Fe XI structure east and south of point '3' (the coronal ``moss", Berger et al. 1999) appears to surround the Ca XVII loops and could be the footpoints of these Ca XVII loops. In fact, many locations with bright Fe XI emission seem to be void of the Ca XVII emission, and vice versa. This indicates that these loops have distinct temperature structures among them. Many of them have a narrow differential emission measure (DEM) distribution along the line of sight that is concentrated either at around 1 million degrees, or at a higher temperature such as 6 million degrees, but not both. See Section 5.1 for more comparisons. 

The XRT image is strikingly similar to that of Ca XVII. The response function of the Be\_med filter is peaked at 10 million degrees and its value falls below 10\% of the peak value at temperatures lower than 3 million degrees (Fig.6). Therefore it is suitable to make direct comparison with the Ca XVII image. The similarity between the two images gives strong proof that our extraction procedure to obtain the Ca XVII emission is correct. 

The structure of the Fe XVI emission shows certain similarities with both Fe XI and Ca XVII. This is not surprising given that the formation temperature of Fe XVI lies in between that of Fe XI and Ca XVII. Note that the Fe XVI emission appears largely diffuse and uniform, especially outside of the core of the active region where many extended, distinct Fe XI loops are seen. This indicates that the plasma at $\sim$3 million degrees is more uniformly distributed across the active region, while the 1 million and 6 million degree plasmas have more localized structures. 
  
We have applied this same procedure to five more raster datasets. The results are shown in Figures 7-11, along with the XRT and Fe XVI images when available. Below we discuss the Ca XVII structure for each of them.
 
\subsection {December 15, 2007, AR10978}
 
This same active region was observed twice on December 15, one from 00:13:49 UT to 05:32:40 UT (`Observation 1') and the other from 18:15:49 UT to 23:34:40 UT (`Observation 2'). The results are shown in Figures 7 and 8. This active region on this day shows obvious differences in morphology from that on December 11. The Ca XVII emission from Observation 1 (Fig.7) exhibits a large extended loop toward the south which resembles the shape of post-flare loops. This active region had been producing many B- and C-class X-ray flares since Dec.11, which included 7 flares on Dec.14 alone. The only C-class event that occurred in this active region within one day of both EIS observations was a C1.1 X-ray flare on 14:11 UT, December 14. These flares are possible causes of this extended Ca XVII loop feature. It is interesting to note that these extended Ca XVII loops (X between 600" to 750", Y between -200" to -300") enclose the smaller, cooler Fe XI loops (which appears mostly dark below Y=-250" for the same X range, see also Section 5.1), which also agree with the standard picture of post-flare loops. Such loop structure in the Fe XVI emission is visible but not as obvious. Several bright loops and some diffuse emission in Ca XVII can be seen at the heart of the active region, but only faint Ca XVII emission, if any, existed around the outer region where many Fe XI and Fe XVI l emissions were present. The XRT image (Be\_med filter) again shows a very good match with the Ca XVII emission. The XRT movie taken during that time shows that the X-ray emission is very dynamic with various loops brightening and fading. For example, the `$\Omega$'-shaped loop at the top was a transient feature that was seen both in X-rays and Ca XVII. 

Observation 2 (Fig.8) was taken 18 hours later. The large post-flare loops in Ca XVII have disappeared (note that there was only one X-ray flare, a B3.8, from this active region in between the two observations) and the diffuse emission at the core of the active region is more obvious. This is the same for the XRT image. To examine more quantitatively the change in line intensity between the two time periods, we calculated the mean values of Fe XI and Ca XVII fluxes in 100 square-pixel areas. We found that the change in Fe XI $\lambda$188.216 intensity is actually quite small-- less than a factor of 5 around the active region core (Y between -100" and -200") and less than a factor of 10 elsewhere. The most significant change occurred at where the Ca XVII post-flare loops were, i.e. the Fe XI emission was depleted where and when these Ca XVII loops were prominent during Observation 1 but these depletions were `filled in' by the time of observation 2 (i.e. its intensity became at about the same level as the surroundings). This does not seem to be due to the slight change in orientation between the two observations due to solar rotation. Interestingly, the Ca XVII emission increased by a factor of 20-200 nearly across the entire region except at where those post-flare loops were. Those Ca XVII loops were around a factor of 100 brighter than its surroundings during Observation 1 but became about the same brightness as the surroundings at Observation 2.  We also note that this post-flare loop feature can be seen in Fe XVI at roughly the same location. All these suggest that the entire region became brighter from Observation 1 to Observation 2 with more increase in Ca XVII than in Fe XI, and the post-flare loops seen in Ca XVII during Observation 1 may have cooled and become part of the 1-3 MK corona. 

\subsection {December 02, 2006, AR10926}

Active region 10926 was observed on December 02, 2006 from 14:06:32 UT to 14:55:45 UT (Figure 9). The dynamical properties of this active region have been presented by Doschek et al. (2007b). The Ca XVII structure is similar to that of the December 11, 2007 data. That is, it is confined to the core of the active region and is composed of several loops with both `fat' and diffuse appearances. There are three places where the loops are particularly bright. It is not clear if they are related to the two C1 X-ray flares that occurred earlier that day at 00:18 and 07:00 UT (there was one B2 flare in between), or these loops are part of the quiescent active region structure. There is no XRT observations during this period. The Fe XVI emission has the same characteristics as AR 10978: bright, localized loops like those of Ca XVII, and diffuse, uniform emission occupying the same space with the Fe XI loops.
 
\subsection {December 17, 2006, AR10930}

Active region 10930 was observed on December 17, 2006 from 16:15:27 UT to 18:29:39 UT at the west limb (Figure 10). A C2.0 X-ray flare occurred at 14:47 UT and the post-flare loops were prominent in all observed lines. In particular, the high temperature Ca XVII loop shows a cusp shape at the loop top and is situated above the cooler loops (e.g. Fe XI emission, also see Section 5.1), consistent with the standard flare model. The readers are referred to Hara et al. (2008) for detailed EIS data analysis for this limb flare event. In this work, the careful extraction of the Ca XVII line out of the blend enables us to see a more realistic structure of the Ca XVII emission. We see that, besides the prominent `fat' loops and the loop-top cusp, the post-flare loops are embedded in an extended, diffuse emission of $\sim$6 million degrees. There is no footpoint brightening in either Ca XVII or O V. The transition region structure on the solar disk is nicely seen in the O V image.
 
\subsection {February 02, 2007, AR10940}
 
The raster observation of this active region was performed from 10:42:12 UT to 11:52:37 UT (Figure 11). The temperature and density structures of this active region have been investigated by Doschek et al. (2007a), and the coronal ``moss" in this active region has been discussed by Warren et al. (2008a). The Ca XVII emission existed only at the northwest part of the region in the form of several loops, and it is much fainter than in other active regions studied here. This high temperature loop structure is similar to the X-ray (Al\_poly) and Fe XVI images shown in Fig.11, as well as the Fe XIV and Fe XV raster images shown in Doschek et al. (2007a). But unlike Ca XVII, these images also exhibit brighter emission at lower temperature (e.g. around 2-3 million degrees) at other parts of the active region. As in other cases, most of these Ca XVII loops are situated at where the Fe XI emission is fainter (see Section 5.1). Similar to AR 10978 (Dec.11, 2007 data), the ``moss" structures (Warren et al. 2008a) lie approximately along the sides of the Ca XVII loops (see also Fig.12), indicating that they are associated with the footpoints of these hot Ca XVII loops. Also interesting is that the O V structure in general follows the morphology of Fe XI. We checked the line profiles of the blend as well as the fitting results, and conclude that the O V structure displayed here is real. Therefore part of this active region contains a particularly cool component. The H$_\alpha$ image (e.g. from http://mlso.hao.ucar.edu) of this active region shows a distinct L-shaped filament at its east and south sides that roughly follows the shape of these Fe XI/O V loops. It is likely that this filament is related to the O V loops seen here.

\section{6 Million Degree vs 1 Million Degree Corona}

\subsection {Spatial Structure and Morphology}

To show more clearly the relative location between the hot Ca XVII and the cool Fe XI emissions, Figure 12 shows the raster images of Fe XI $\lambda$188.216 (obtained from the 2-Gaussian fits) overplotted with the contours of Ca XVII $\lambda$192.858 fluxes for the six data sets. Similar to the Ca XVII images shown above, those pixels with Ca XVII flux less than 10\% of the blend are not taken into account. The coronal moss structures in the Feb.02, 2007 and Dec.11, 2007 data are  indicated, showing that they generally surround the bright Ca XVII loops. We can see that most locations with strongest Ca XVII emission are where the Fe XI emission is relatively weak. This is especially clear for the Dec.02, 2006 (upper left), Feb.02, 2007 (lower left) and Dec.11, 2007 (upper right) data. It is not so clear in the Dec.15, 2007 data (both observations) probably because the AR was observed at a more slanted view angle. As mentioned in Section 4.1, this implies a narrow DEM distribution {\it along the line of sight} that is concentrated either at around 1 million degrees, or at a higher temperature such as 6 million degrees, but not both. Furthermore, judging from that the Fe XVI emission (with peak formation temperature at 2.5 million degrees) co-exists with both the Fe XI and Ca XVII emissions, we can imagine that the DEM distribiution at an active region loop probably has a FWHM of around 2 million degrees, not much narrower and not much wider, and the peak temperature at one given moment is determined by the history of heating and cooling of the loop (cp. Fig.6). We need to emphasize that a given line emission at a given location is a convolution of its emissivity with the DEM distribution along the line of sight. {\it A detailed DEM analysis is  necessary} to demonstrate if the above inference is indeed true.

The spatial morphology shown would have important implications on how an active region corona is heated and cooled. It is likely that only a fraction of the AR loops (in a non-flaring AR) is heated to very high temperature ($\sim$6 MK) at one given moment, and these high temperature loops are obviously tied to regions with strongest magnetic field strengths. This is not surprising since it is commonly acknowledged that the coronal heating rate is related to the photosheric magnetic field strength by some power law (e.g. Yashiro \& Shibata 2001, van Driel-Gesztelyi et al. 2003, Schrijver et al. 2004). Do these hot loops subsequently cool, or are they repeatedly heated and never cool significantly? Are those extended 1 MK loops (or even 3 MK loops) all remnants of the once-hot loops, or are some of them never heated to 6 MK at all? High time cadence observations to look for time variations from spectral lines covering a wide range of temperatures and in a variety of loops would shed lights on these questions (e.g. Mariska et al. 2007). The Ca XVII line provides essential information for the high-temperature end of the active region corona and is a useful and viable spectral line for understanding the thermal structure and heating processes in active regions (Parenti et al. 2006; Patsourakos \& Klimchuk 2007).

\subsection {Emission Measure}

EIS observations of active regions generally contain many lines that allow a DEM analysis to probe the temperature structure, and contain certain line pairs that are suitable for density diagnostics (Young et al. 2007,2009). As mentioned in Section 1, Ca XVII is the most practical line to extend the temperature analysis of non-flaring active regions to its high end. We will present such DEM analysis in a separate paper. In this paper, we only explore the temperature structure using Fe XI $\lambda$188.216 and Ca XVII $\lambda$192.858 lines as proxies of the 1 million and 6 million degree corona. That is, we calculate the ratio of the emission measure at 6 MK to that at 1 MK using the emissivities at the peak formation temperatures of the two lines:
\begin{equation}
{EM_{6MK}\over EM_{1MK}} = {I_{CaXVII}\over I_{FeXI}} {{A_{Fe}/A_{Fe,ph}}\over {A_{Ca}/A_{Ca,ph}}} {\varepsilon_{FeXI}(T_e=1.26MK)\over \varepsilon_{CaXVII}(T_e=6.3MK)}
\end{equation}
where $I$ is the line flux, $A_{el}/A_{el,ph}$ is the elemental abundance relative to its photospheric value and $\varepsilon(T_e)$ is the emissivity (or contribution function) which is a function of the electron temperature and density. In this simple calculation, we assume an electron density of $10^{10}$ cm$^{-3}$, therefore $\varepsilon_{FeXI}(T_e=1.26MK)/ \varepsilon_{CaXVII}(T_e=6.3MK)=17.1$. This assumption of the electron density is reasonable for active regions, especially for bright features (e.g. Doschek et al. 2007a, Young et al. 2009). The emissivity of Ca XVII $\lambda$192.858 is insensitive to the electron density, and the emissivity of Fe XI $\lambda$188.216 increases by $\sim$30\% if the density is $10^{9}$ cm$^{-3}$ instead. We assume $A_{Fe}/A_{Fe,ph}=A_{Ca}/A_{Ca,ph}$ given that both are low First-Ionization-Potential (FIP) elements so the ratio should not be far off from 1 whether the FIP effect is present or not. The emissivities are obtained from the CHIANTI database version 5.2 (Dere et al. 1997; Landi et al. 2006). Specifically, readers are referred to Tachiev \& Froese Fischer (1999) for O V lines, Young (1998) for Fe XI lines, and Zhang \& Sampson (1992) for the Ca XVII line. We adopt the ionization equilibria from the updated compilation of Bryans et al. (2006). The photospheric abundances are adopted from Grevesse \& Sauval (1998).

Figure 13 shows the results for the 5 AR data sets (excluding the limb flare data on Dec.17, 2006). The left panels show the probability distribution function (PDF) plots for Fe XI $\lambda$188.216 (dashed line) and Ca XVII (solid line) fluxes. Right panels are the PDF plots for $EM_{6MK}/EM_{1MK}$. All pixels in the raster images shown in Figs.4,7-11 for individual active regions are included, but those with Ca XVII flux less than 10\% of the blend are not taken into account in the PDF. As a result, the PDF plots for log$_{10} EM_{6MK}/EM_{1MK}$ have a cutoff at around $-0.35$ (i.e. log$_{10}(17.1*0.26*0.1)$ where $17.1*0.26$ is the emissivity ratio of $\varepsilon_{FeXI \lambda 192.813}(T_e=1.26MK)/ \varepsilon_{CaXVII}\lambda 192.858(T_e=6.3MK)$, 0.1 is the 10\% cutoff point). We see that: 1) The intensity of the Fe XI emission has a narrow distribution mostly within a factor of 10, even though the most-probable intensity can be different. This is true for all ARs studied here. This indicates that the mechanisms for producing the 1MK corona are not much different for different ARs. 2) The intensity of the Ca XVII emission has a wider distribution which may be partly due to the uncertainty of the extraction procedure. It seems to consist of two components, a `bright' (high intensity `wing') component from those ultra-bright Ca XVII loops that is different for different ARs reflecting subtle differences in coronal heating, and a main component which is probably from the diffuse Ca XVII emission and, like Fe XI, is similar for all ARs. 3) $EM_{6MK}/EM_{1MK}$ can be as high as 10 at the core of the active regions. Outside of the active region core where the 1 million degree loops are abundant and the Ca XVII emission is relatively week, we can put an upper limit on $EM_{6MK}/EM_{1MK}$ to be about 0.5. Below this upper limit, the Ca XVII emission can not be extracted from the blend with confidence. This information, as well as the morphology and spatial structures discussed in previous sections, can serve as empirical constraints for coronal heating models (e.g. Schrijver et al. 2004; Parenti et al. 2006; Warren \& Winebarger 2007; Klimchuk et al. 2008). This study implies that at the core of some active regions, the heating needs to be high enough to produce emission measure at 6 million degrees 1-10 times higher than that at 1 million degrees. On the other hand, outside of the active region core where long, 1 million degree loops are abundant, the heating should be low enough so that $EM_{6MK}/EM_{1MK}$ does not exceed $\sim$0.5.

As a final note, one may wonder if these Ca XVII emissions indeed come from a plasma at around 6 MK, and not from a lower-temperature plasma, say 3 MK, where its emission measure happens to be large enough to produce the observed Ca XVII emission. Since these datasets do not contain Ca XV or Ca XVI lines (Warren et al. 2008b) for us to derive temperature from Ca line ratios, we use the Fe XVI line (Fig.6) instead and look into this question in two approaches. We use points 1 and 2 in Fig.5 for the Dec.11, 2007 data as examples. The two locations have $EM_{6MK}/EM_{1MK}$ values of 3.9 and 11.8 for pt.1 and 2, respectively. First, we use the Ca XVII to Fe XVI intensity ratio to derive the electron temperature (cp.Eq.(1)). The underlying assumptions are that both lines emit solely from that same temperature and their abundances are the same. We obtain $T_e$ to be $4.07\times 10^6$ K and $4.17\times 10^6$ K, and the emission measure to be $1.1\times 10^{29}$ cm$^{-5}$ and $1.4\times 10^{29}$ cm$^{-5}$ (assuming photospheric abundance) for pt.1 and 2, respectively. 
We see that these temperatures are in between the Ca XVII and Fe XVI formation temperatures (Fig.6). We also check one of the brightest Ca XVII emission in this dataset and obtain a $T_e$ of $4.85\times 10^6$ K. The second approach is to start from the emission measure. If we assume that the Fe XVI emission solely comes from a plasma at 3 MK, the emission measure would be $7.4\times 10^{28}$ cm$^{-5}$ and $9.1\times 10^{28}$ cm$^{-5}$ for pt.1 and 2, respectively. If there is no plasma above this temperature, such emission measures would produce only 2.9\% and 2.3\% of the observed Ca XVII emission. Similarly, the emission measure at 6 MK, if it is the only source for the observed Ca XVII emission, would be too small to be also the only source for the observed Fe XVI emission. Both approaches are actually two faces of the same conclusion: the temperature structure is probably multi-thermal (along the line of sight), and the observed Ca XVII emission can at least partly come from a plasma at 6 MK. As emphasized throughout this paper, the appropriate way to investigate temperature structure is to perform a DEM analysis. Such efforts have been underway and we do see that the EM at 6 MK can be comparable to that at 1 MK at certain locations where the Ca XVII emission is strong. Such results will be presented in a separate paper.

\section{Summary}
 
We have developed a specific procedure to extract the Ca XVII $\lambda$192.858 line from the blending with two Fe XI and six O V lines in the Hinode/EIS data. We have performed this procedure on the raster data of five active regions and a limb flare, and demonstrated that the Ca XVII line can be satisfactorily extracted from the blend if the Ca XVII flux contributes to at least $\sim$10\% of the blend. This Ca XVII line can thus be used to probe the thermal structure of the active region corona at its high temperature end and provide valuable constraints for coronal heating models. We believe this procedure extracts the Ca XVII emission more accurately from the blend, compared to those by subtracting the contribution of Fe XI and O V based on other observed Fe XI and O V lines. The latter may be satisfactory when Ca XVII emission is a major fraction of the blend but the error can be much larger when Ca XVII is faint or even comparable to other lines in the blend. To perform detailed DEM analysis, this procedure is preferable so to obtain more accurate information about the thermal structure near 6 million degrees. The main results of this study are summarized below.
 
\noindent{{1) The Ca XVII emission is strongest around the active region core and appears as fat, low-lying loops coexisting with some weaker, diffuse emission surrounding those bright Ca XVII loops. This implies that the active region corona can be largely heated to very high temperature ($>$ 5 million degrees) only where the photospheric magnetic fields are the strongest.}

\noindent{{2) The striking similarity in morphology between the Ca XVII emission and the X-ray emission from XRT indicates that our procedure for extracting the Ca XVII out of the blend is accurate, and the spatial structure across a non-flaring active region is very similar in the range of 5-10 million degrees. On the other hand, the different spatial structure/morphology exhibited among Ca XVII, Fe XVI and Fe XI lines indicates that the DEM distribution below $\sim$6 MK at any given location in an active region probably has a FWHM of around 2 million degrees, not much narrower and not much wider. Detailed DEM analyses are required to unambiguously differentiate the thermal structure and evolution among spatial structures.}


\noindent{{3) The emission measure ratio of the 6 million degree plasma relative to the cooler 1 million degree plasma in the core of the active regions, using the Ca XVII to Fe XI line intensity ratio as a proxy, can be as high as 10. Outside of the active region core where the 1 million degree loops are abundant, this study places an upper limit of about 0.5 for the ratio. This information, as well as the morphology and spatial structures discussed here, provide empirical constraints for coronal heating models.} 

The authors acknowledge support from the NASA {\it Hinode} program, and the NRL/ONR 6.1 basic research program. Y.-K. K. thanks M. Weber for the help with the XRT temperature response functions. We thank the anonymous referee for valuable comments that greatly improve this paper.
{\it Hinode} is a Japanese mission developed and launched by
ISAS/JAXA, collaborating with NAOJ as domestic partner, and NASA (USA)
and STFC (UK) as international partners.  Scientific operation of the
{\it Hinode} mission is conducted by the {\it Hinode} science team
organized at ISAS/JAXA.  This team mainly consists of scientists from
institutes in the partner countries.  Support for the post-launch
operation is provided by JAXA and NAOJ, STFC, NASA, ESA (European
Space Agency), and NSC (Norway).  We are grateful to the {\it Hinode}
team for all their efforts in the design, build, and operation of the
mission. MPFIT is a public-domain IDL program developed by C. B. Markwardt (http://www.physics.wisc.edu/\~{}craigm/idl/fitting.html).

\clearpage

\begin{table}\label{table1}
\caption{Spectral Lines in the $\lambda$188 and $\lambda$192 Blend}
\begin{center}
\begin{tabular}{lllll}
\hline
\hline
Ion & Transition & Wavelength\tablenotemark{a} (\AA) & Relative Strength\tablenotemark{b} & log$_{10}$T$_{max}$\tablenotemark{c}\\
\hline
\ion{Fe}{11} & 3s$^2$3p$^4$ $^3$P$_2$ - 3s$^2$3p$^3$3d $^3$P$_2$ &
188.216 & 1.0 & 6.1\\
\ion{Fe}{11} & 3s$^2$3p$^4$ $^3$P$_2$ - 3s$^2$3p$^3$3d $^1$P$_1$ &
188.299 & 0.70& 6.1\\
\ion{Fe}{11} & 3s$^2$3p$^4$ $^3$P$_1$ - 3s$^2$3p$^3$3d $^3$P$_2$ &
192.813 & 0.26& 6.1\\
\ion{Fe}{11} & 3s$^2$3p$^4$ $^3$P$_1$ - 3s$^2$3p$^3$3d $^1$P$_1$ &
192.901 & 0.0052 & 6.1\\
&&&&\\
\ion{O}{5} & 2s2p $^3$P$_0$ -  2s3d $^3$D$_1$ &
192.750 & 0.1729 & 5.4\\
\ion{O}{5} & 2s2p $^3$P$_1$ -  2s3d $^3$D$_2$ &
192.797 & 0.3193 & 5.4\\
\ion{O}{5} & 2s2p $^3$P$_1$ -  2s3d $^3$D$_1$ &
192.801 & 0.1296 & 5.4\\
\ion{O}{5} & 2s2p $^3$P$_2$ -  2s3d $^3$D$_3$ &
192.904 & 1.0 & 5.4\\
\ion{O}{5} & 2s2p $^3$P$_2$ -  2s3d $^3$D$_2$ &
192.911 & 0.1063 & 5.4\\
\ion{O}{5} & 2s2p $^3$P$_2$ -  2s3d $^3$D$_1$ &
192.915 & 0.00863 & 5.4\\
&&&&\\
\ion{Ca}{17} & 2s$^2$ $^1$S$_0$ -  2s2p $^1$P$_1$ &
192.858 & -- & 6.8\\
\hline
\end{tabular}
\tablenotetext{a}{O V from Fuhr et al. (1999). For Fe XI, see Young (1998) and Brown et al. (2008). For Ca XVII, see Brown et al. (2008) (see also Dere 1978).}
\tablenotetext{b}{For Fe XI, the ratios are relative to the 188.216 \AA\ line and are empirically determined (see Section 3.1). For O V, the ratios are relative to the strongest 192.904 \AA\ line and are adopted from CHIANTI database v5.2.}
\tablenotetext{c}{Peak formation temperature (in K) from CHIANTI database v5.2 with ionization equilibrium from Bryans et al. (2006).}
\end{center}
\end{table}

\clearpage

\begin{figure}
\centerline{%
\includegraphics[width=5.5in]{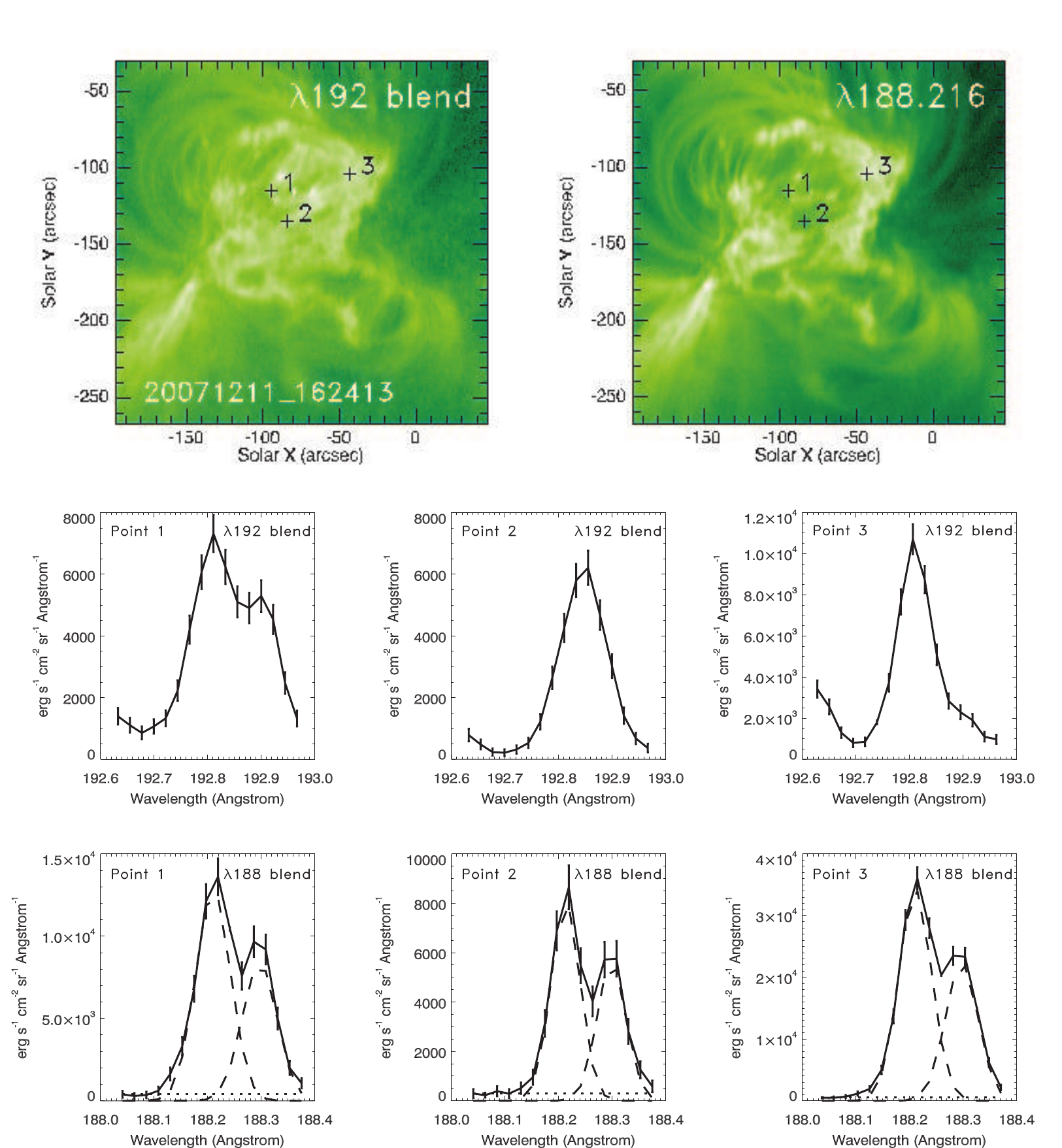}}
\caption{Upper panel: raster images of the $\lambda$192 blend (left) and the Fe XI $\lambda$188.216 line (right). The fluxes of Fe XI $\lambda$188.216 are obtained by fitting the $\lambda$188 blend with two Gaussians. The fluxes of the $\lambda$192 blend are obtained by summing across the profile of the entire blend that includes the Ca XVII line, 2 Fe XI lines and 6 O V lines with background level subtracted.  The raster observations were on AR10978 on December 11, 2007 from 16:24:13 UT to 17:35:04 UT. The six plots below the two raster images are the spectra at the three marked locations for the $\lambda$192 blend (middle panels) and the Fe XI $\lambda$188 blend (lower panels). The results of the 2-Gaussian fitting of the Fe XI $\lambda$188 blend are also shown (dashed lines: the two Fe XI components, dotted line: constant background level).}
\label{fig1}
\end{figure}

\begin{figure*}
\centerline{%
\includegraphics[width=6in]{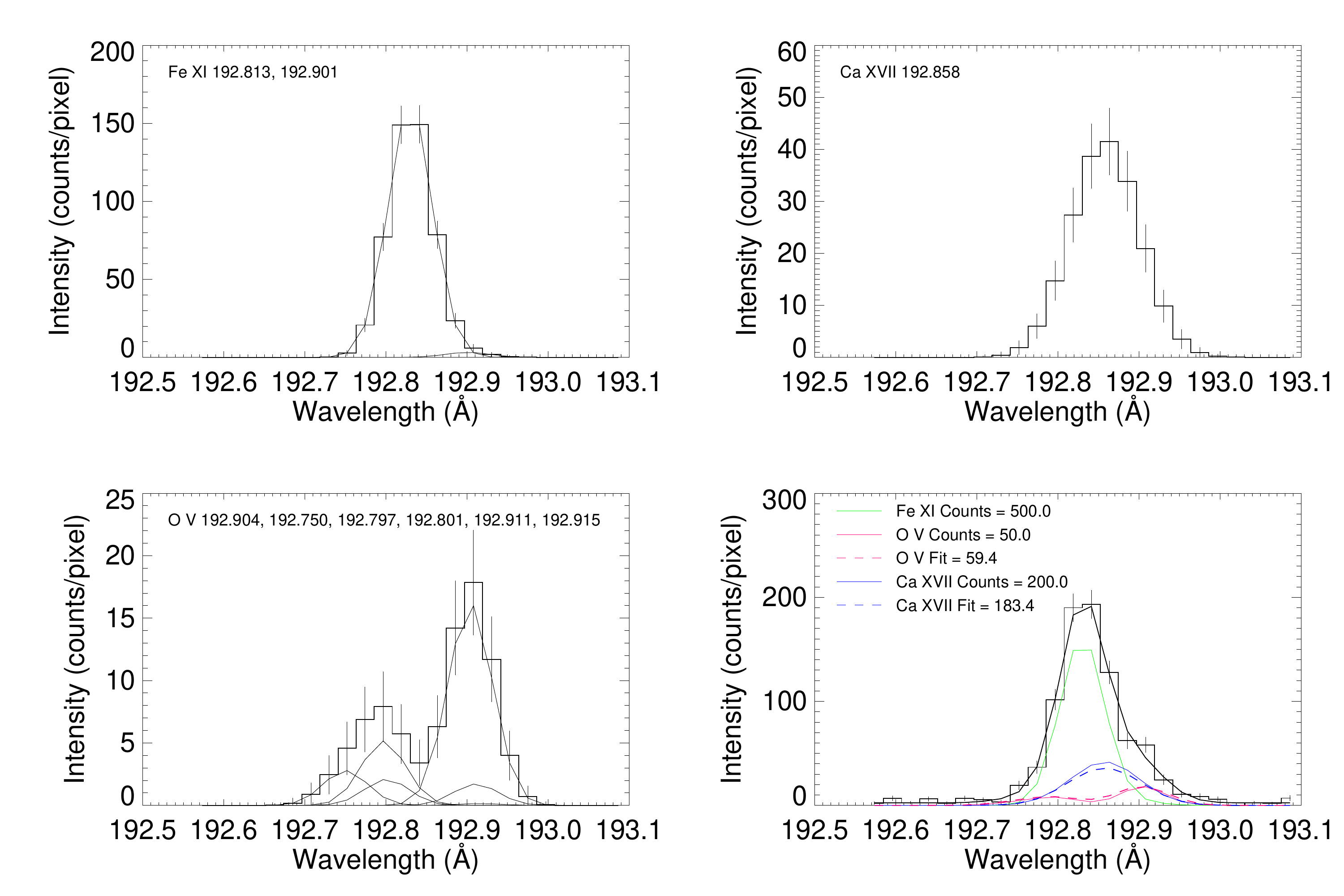}}
\caption{An example Monte Carlo simulation of the \ion{Fe}{11},
\ion{Ca}{17}, and
 \ion{O}{5} deconvolution. The first three panels show the individual
components and the
 statistical uncertainties. The final panel shows the composite profile
with noise
 added. In the final panel the input and fitted Gaussians are also shown
for each
 component.}
\label{fig:mc1}
\end{figure*}

\begin{figure}
\centerline{%
\includegraphics[width=4.5in]{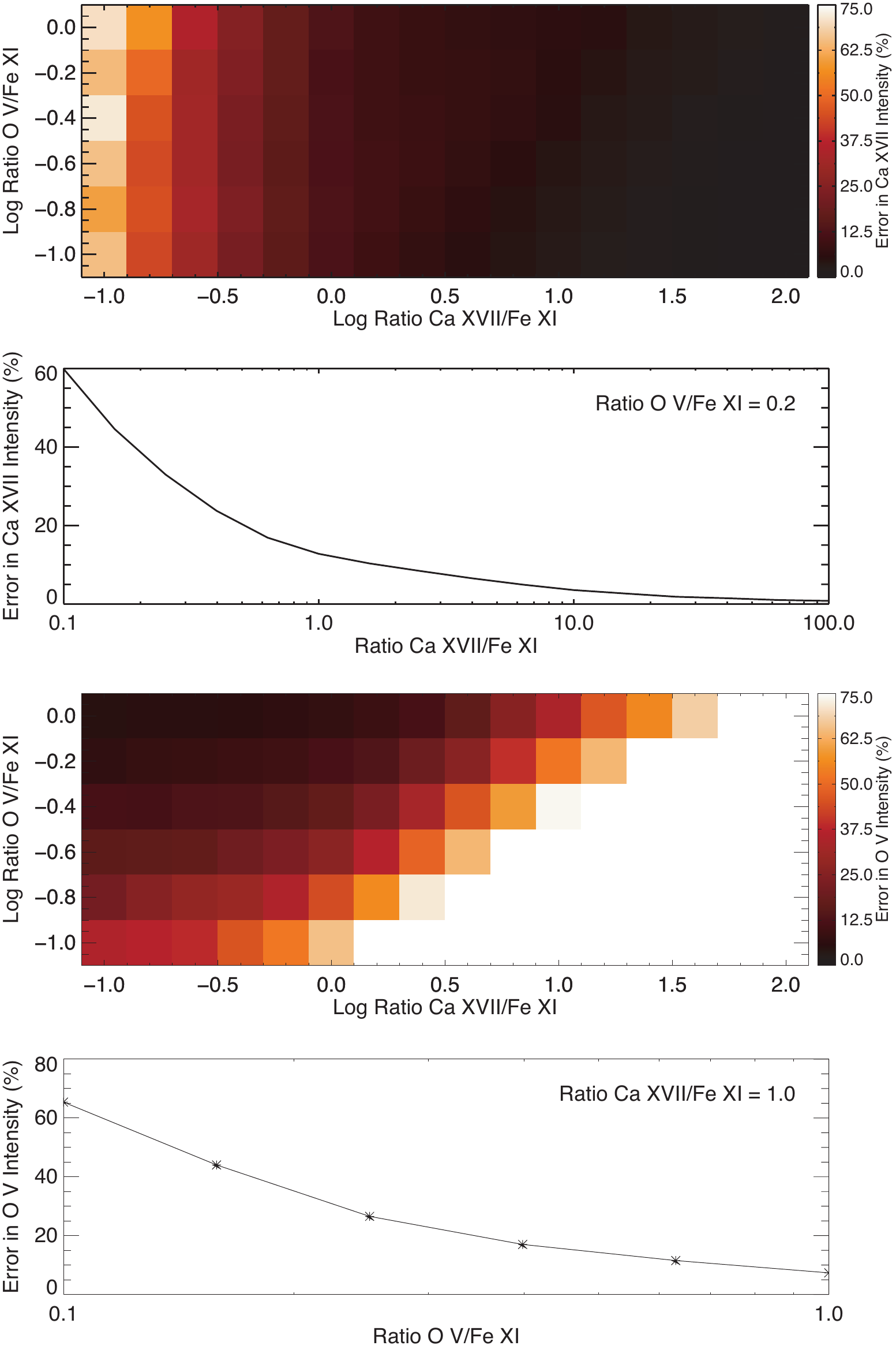}}
\caption{Monte Carlo simulations of the \ion{Fe}{11}, \ion{Ca}{17}, and
\ion{O}{5}
 deconvolution. Top panel: Error in the \ion{Ca}{17} line intensity as a
 function of \ion{O}{5}/\ion{Fe}{11} and \ion{Ca}{17}/\ion{Fe}{11} intensity ratios. 2nd from top: 
Error in the \ion{Ca}{17} line intensity for a fixed \ion{O}{5} contribution. 
3rd from top: Error in the \ion{O}{5} line intensity as a
 function of \ion{Ca}{17}/\ion{Fe}{11} and \ion{O}{5}/\ion{Fe}{11} intensity ratios. Bottom panel: 
Error in the \ion{O}{5} line intensity for a fixed \ion{Ca}{17} contribution. As
expected, the uncertainty in the \ion{Ca}{17} and O V intensities go up as it contributes less
to the blended profile.}
\label{fig:mc2}
\end{figure}

\begin{figure}
\centerline{%
\includegraphics[width=5.5in]{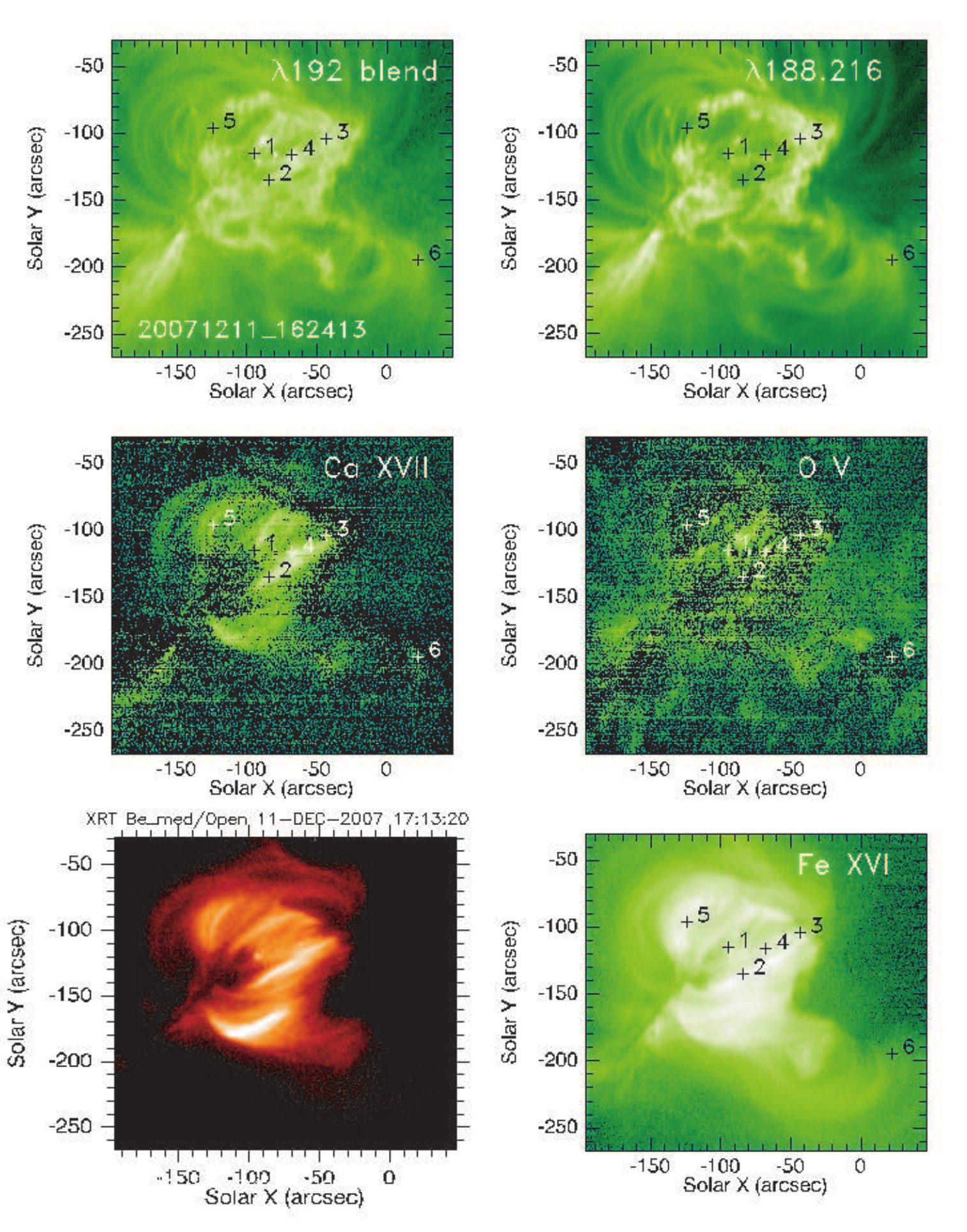}}
\caption{Upper panels: raster images of the $\lambda$192 blend (left) and Fe XI $\lambda$188.216 (right) for AR10978 on December 11, 2007. 
Middle panels: images of Ca XVII $\lambda$192.858 (left) and O V $\lambda$192.904  (right) derived from this work. Six locations are marked to show examples of the fitting results (see Figure 5). Lower left: Image of Hinode/XRT (Be\_med filter) at 17:13:20 UT. Lower right: raster image of Fe XVI $\lambda$262.98.}
\label{fig4}
\end{figure}

\begin{figure}
\centerline{%
\includegraphics[width=6in]{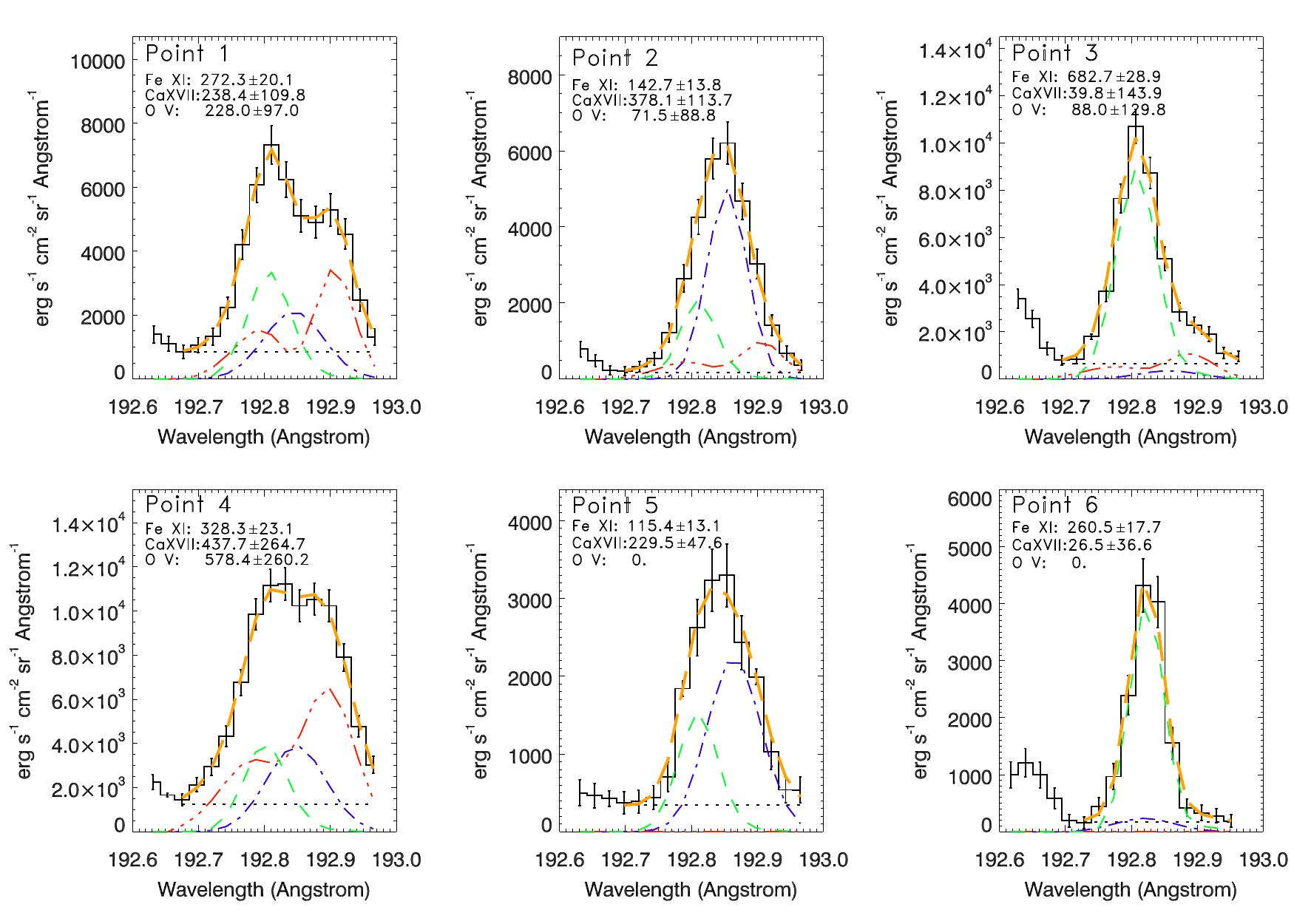}}
\caption{Examples of the fitting results for the December 11, 2007 data (Fig.4). Points 1,2,3 are the same as those in Fig.1. Green dashed line is the two Fe XI components. Blue dash-dot line is the Ca XVII line. Red dash-dot-dot-dot line is the six O V components. The horizontal black dotted line is the background level. Yellow dashed line is the sum of all fitted components. Black solid line with error bars is the data. Points 1 and 4 have strong O V emission. Points 2 and 5 have strong Ca XVII emission. Points 3 and 6 are dominated by the Fe XI emission. The derived fluxes with errors (in erg s$^{-1}$ cm $^{-2}$ sr $^{-1}$) from the multi-Gaussian fitting for Fe XI 192.813 \AA, Ca XVII, and O V 192.904 \AA\ are listed on the plots.}
\label{fig5}
\end{figure}

\begin{figure}
\centerline{%
\includegraphics[width=5.5in]{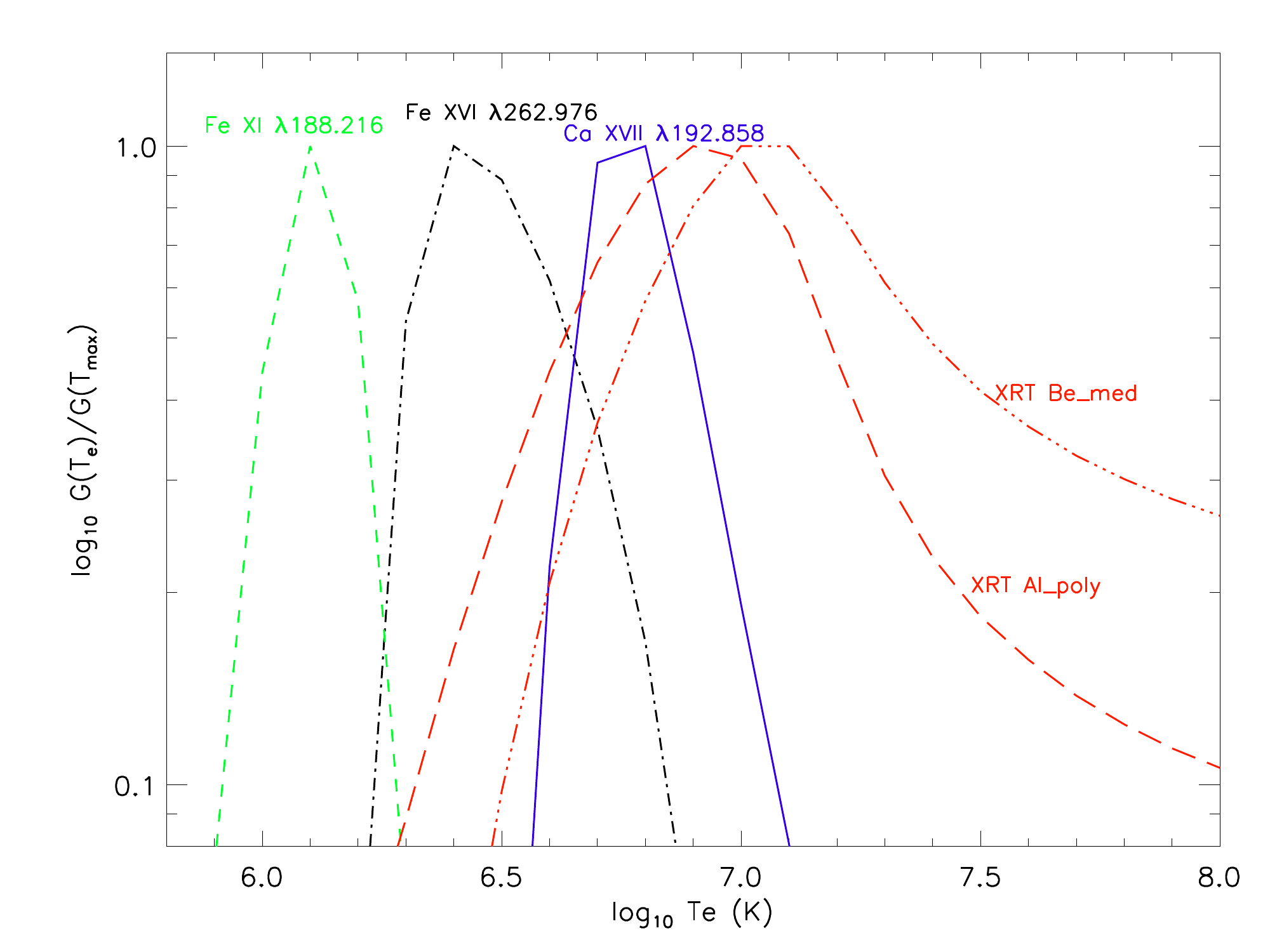}}
\caption{Emissivities (G(T)) of Fe XI $\lambda$188.216, Fe XVI $\lambda$262.976, Ca XVII $\lambda$192.858, and the temperature response function for the Be\_med and Al\_poly filters of Hinode/XRT. The curves are all normalized to their respective maximum. The emissivities are obtained from CHIANTI database v5.2 with ionization equilibrium from Bryans et al. (2006). The XRT effective areas are calculated using the standard calibration routines available in SolarSoft with CCD contamination layer taken into account at Feb.02, 2007, 11:00 UT  (Al\_poly filter) and  Dec.11, 2007, 17:13 UT (Be\_med filter). The default APED model is used to calculate the XRT temperature response function.}
 \label{fig6}
\end{figure}

\begin{figure}
\centerline{%
\includegraphics[width=5.5in]{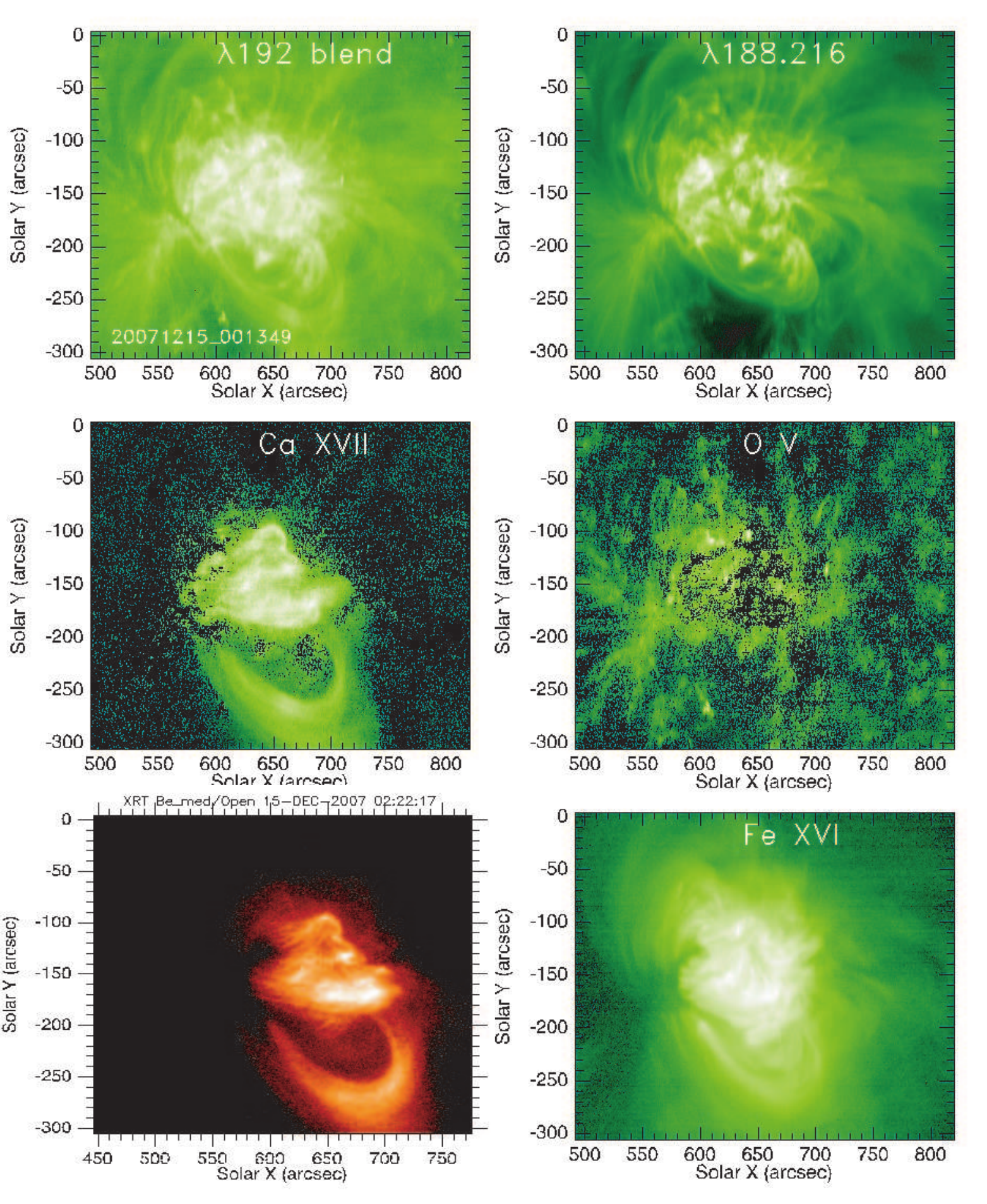}}
\caption{Upper panels: raster images of the $\lambda$192 blend (left) and Fe XI $\lambda$188.216 (right). The raster observations were for AR10978 on December 15, 2007 from 00:13:49 UT to 05:32:40 UT. The image shown here is part of the observations from 00:13:49 UT to 04:23:12 UT.
Middle panels: images of Ca XVII $\lambda$192.858 (left) and O V $\lambda$192.904 (right) derived from this work. Lower left: Image of Hinode/XRT (Be\_med filter) at 02:22:17 UT. Lower right: raster image of Fe XVI $\lambda$262.98.} 
\label{fig7}
\end{figure}

\begin{figure}
\centerline{%
\includegraphics[width=5.5in]{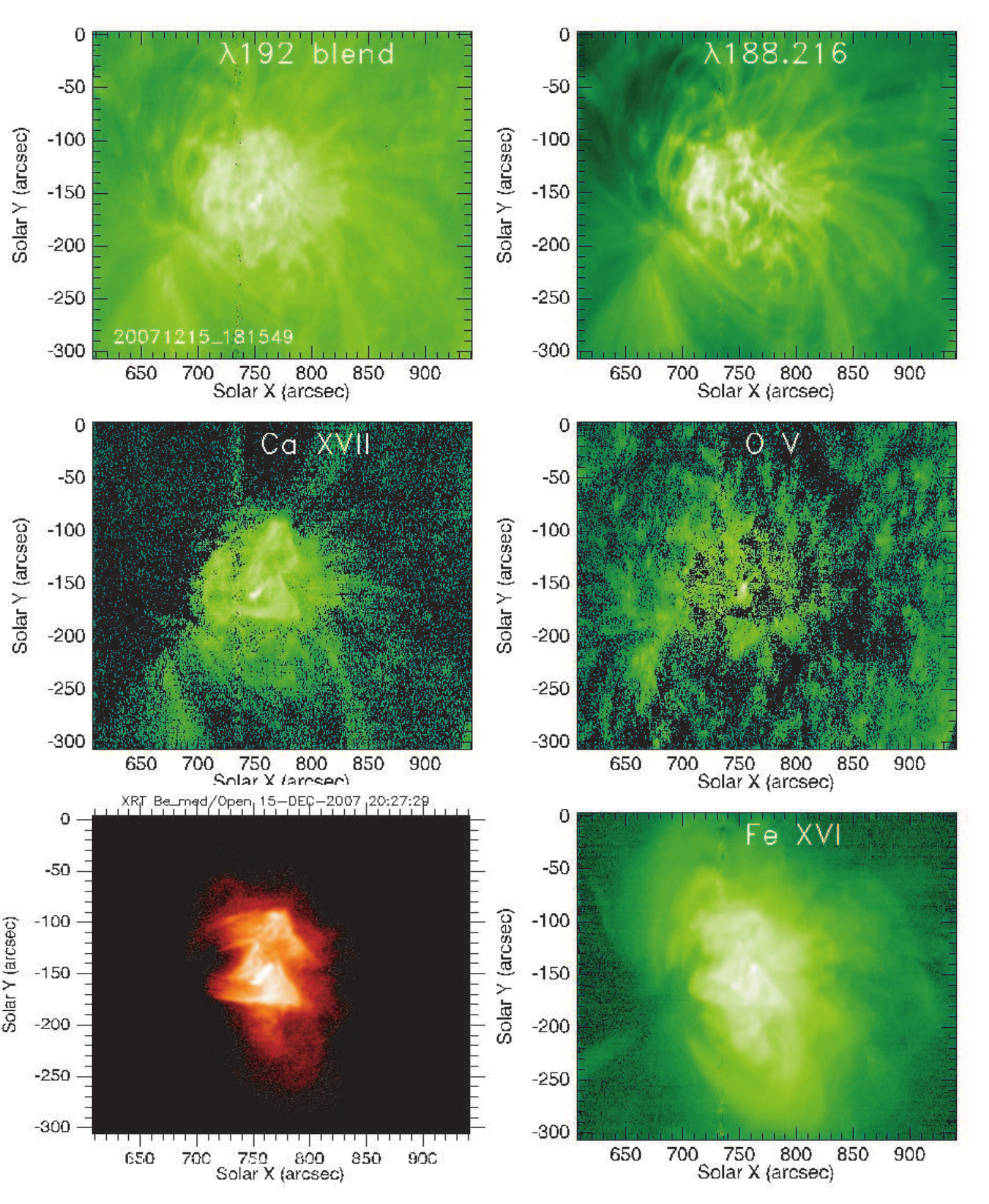}}
\caption{Upper panels: raster images of the $\lambda$192 blend (left) and Fe XI $\lambda$188.216 (right). The raster observations were for AR10978 on December 15, 2007 from 18:15:49 UT to 23:34:40 UT. The image shown here is part of the observations from 18:15:49 UT to 22:25:12 UT.
Middle panels: images of Ca XVII $\lambda$192.858 (left) and O V $\lambda$192.904 (right) derived from this work. Lower left: Image of Hinode/XRT (Be\_med filter) at 20:27:29 UT. Lower right: raster image of Fe XVI $\lambda$262.98.}
\label{fig8}
\end{figure}

\clearpage

\begin{figure}
\centerline{%
\includegraphics[width=5.5in]{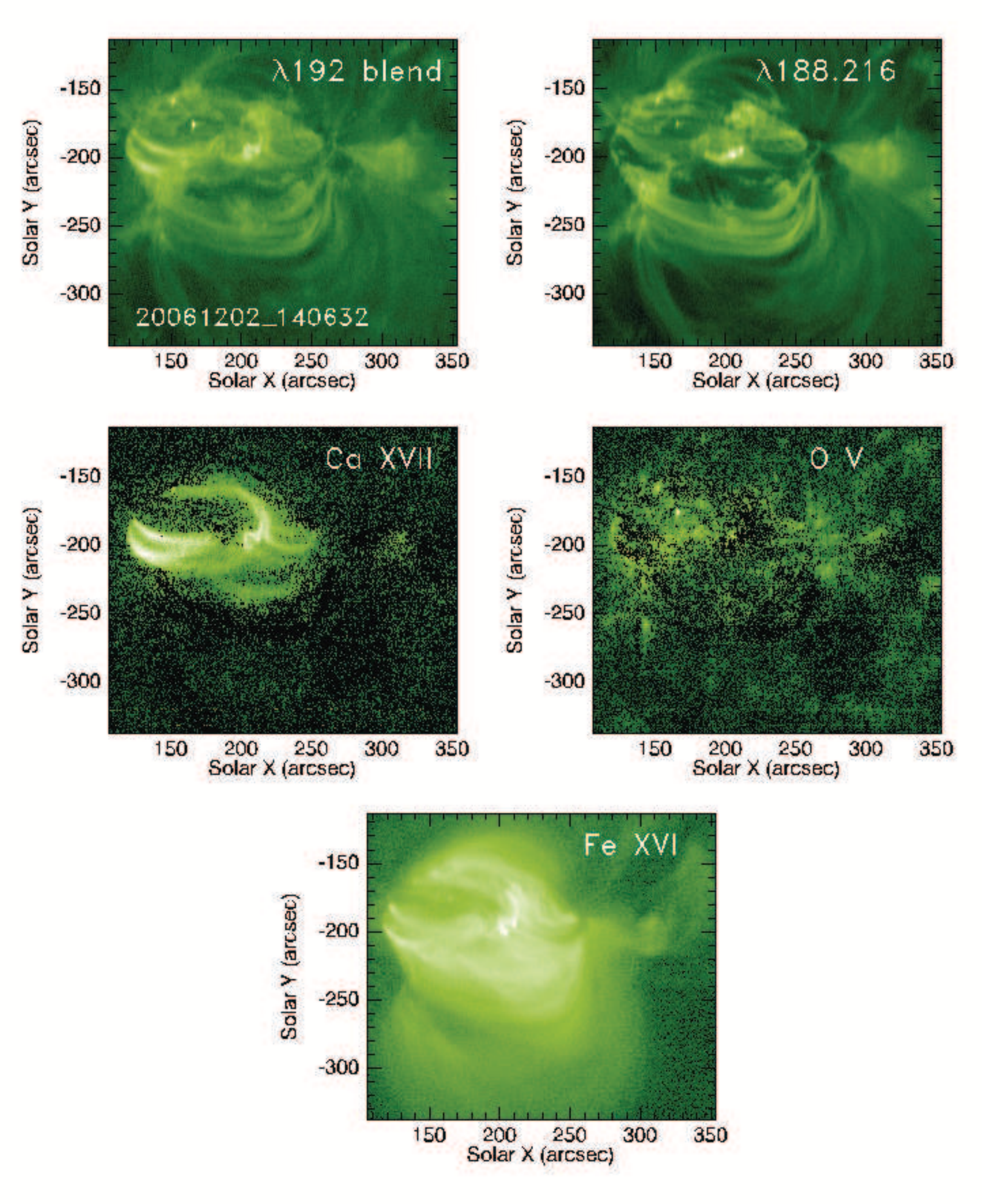}}
\caption{Upper panels: raster images of the $\lambda$192 blend (left) and Fe XI $\lambda$188.216 (right). The raster observations were for AR10926 on December 02, 2006 from 14:06:32 UT to 14:55:45 UT. 
Middle panels: images of Ca XVII $\lambda$192.858 (left) and O V $\lambda$192.904 (right) derived from this work. Lower panel: Raster image of Fe XVI $\lambda$262.98. There is no XRT image available during this EIS raster observations.}
\label{fig9}
\end{figure}

\begin{figure}
\centerline{%
\includegraphics[width=5.5in]{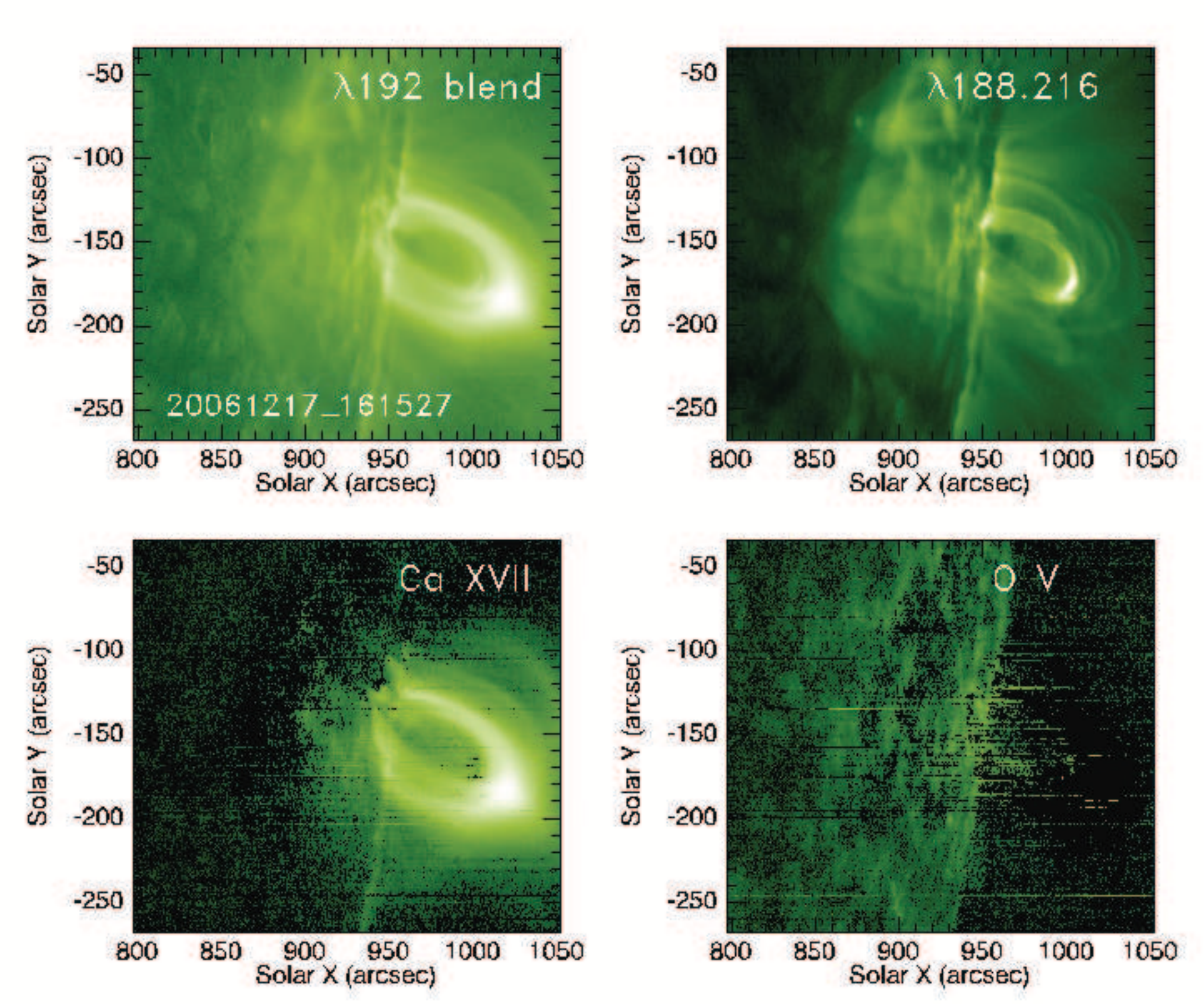}}
\caption{Top panels: raster images of the $\lambda$192 blend (left) and Fe XI $\lambda$188.216 (right). The raster observations were for AR10930 on December 17, 2006 from 16:15:27 UT to 18:29:39 UT. 
Bottom panels: images of Ca XVII $\lambda$192.858 (left) and O V $\lambda$192.904 (right) derived from this work. No XRT data were taken during the raster observations. The Fe XVI $\lambda$262.98 line was not included in this EIS study.}
\label{fig10}
\end{figure}

\begin{figure}
\centerline{%
\includegraphics[width=5.5in]{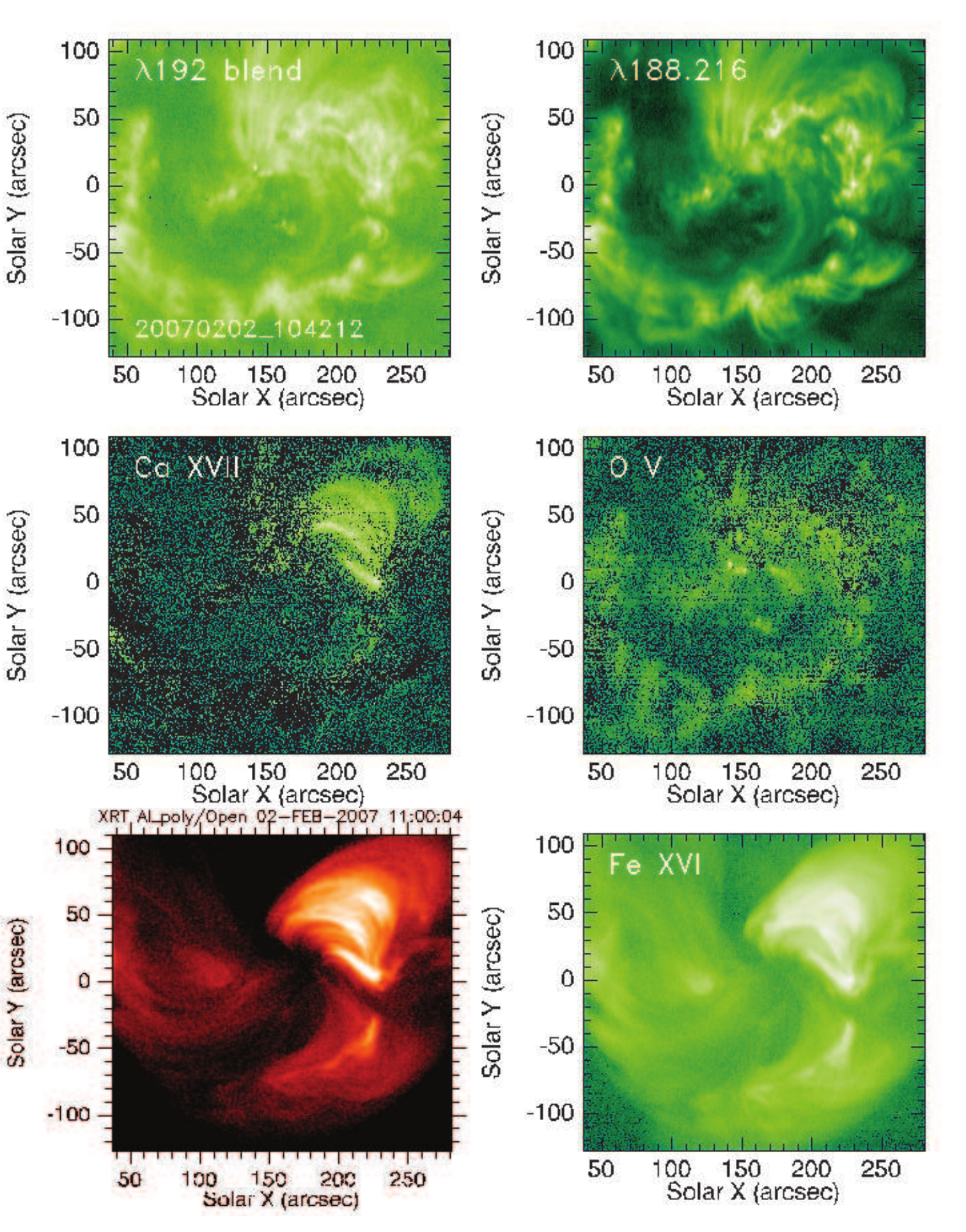}}
\caption{Upper panels: raster images of the $\lambda$192 blend (left) and Fe XI $\lambda$188.216 (right). The raster observations were for AR10940 on February 02, 2007 from 10:42:12 UT to 11:52:37 UT. 
Middle panels: images of Ca XVII $\lambda$192.858 (left) and O V $\lambda$192.904 (right) derived from this work. Lower left: Image of Hinode/XRT (Al\_poly/Open filter) at 11:00:04 UT. Lower right: Raster image of Fe XVI $\lambda$262.98.}
\label{fig11}
\end{figure}

\begin{figure}
\centerline{%
\includegraphics[width=5in]{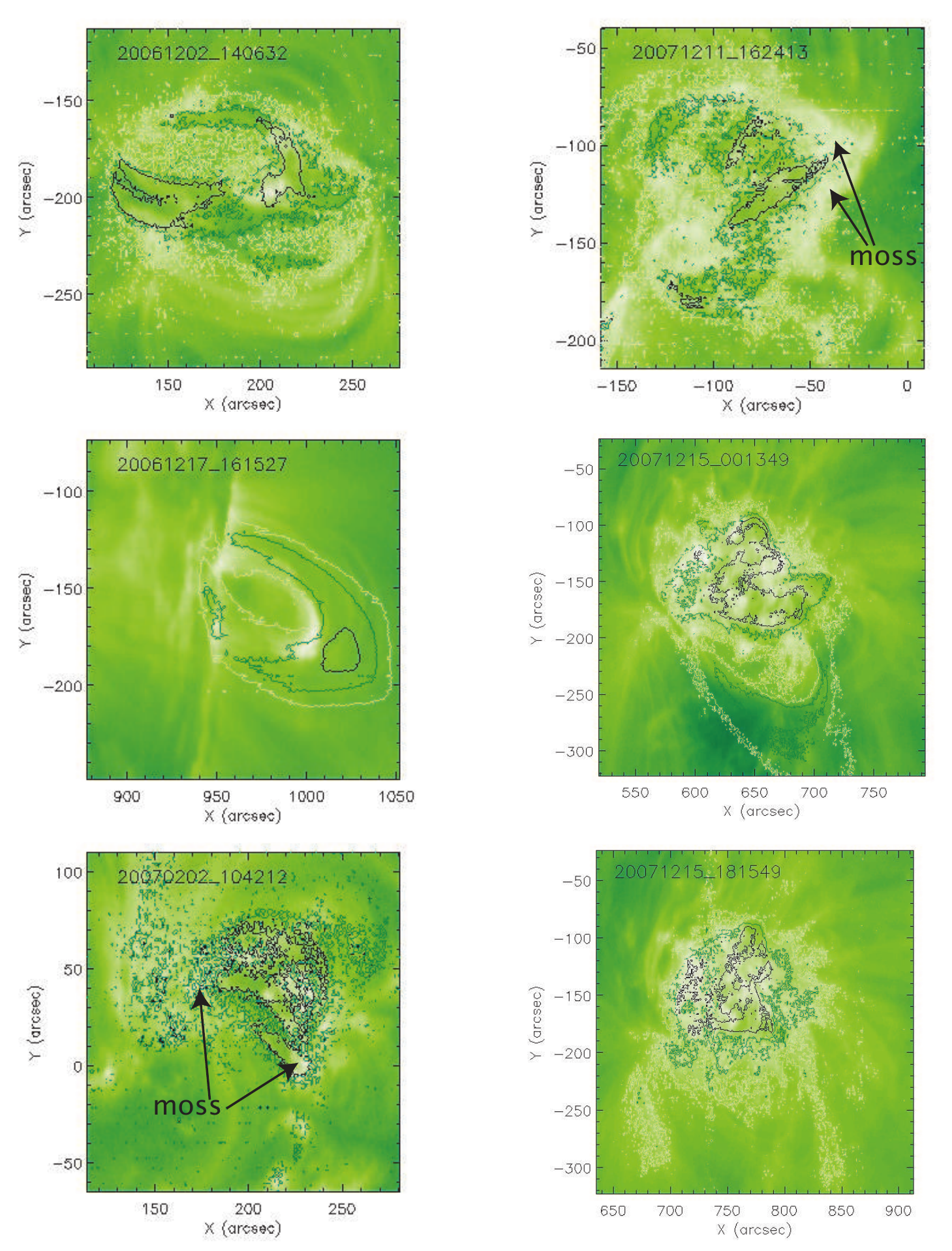}}
\caption{Raster images of  Fe XI $\lambda$188.216 overplotted with contours of Ca XVII $\lambda$192.858 fluxes. Upper left: AR 10926 (Dec.02, 2006), middle left: AR 10930 (Dec.17, 2006), lower left: AR 10940 (Feb.02, 2007) with locations of the coronal moss marked, upper right: AR 10978 (Dec.11, 2007) with locations of the coronal moss marked, middle right: AR 10978 (Dec.15, 2007, observation 1), lower right: AR 10978 (Dec.15, 2007, observation 2). The contour levels (log$_{10}$Flux, all in erg/s/cm$^2$/sr) are: black-- 2.45, green-- 2.05, white--1.65 except for AR 10930 (black-- 3.6, green-- 3.1, white--2.7) and AR 10940 (black-- 1.9, green-- 1.6). }
\label{fig12}
\end{figure}

\begin{figure}
\centerline{%
\includegraphics[width=3in]{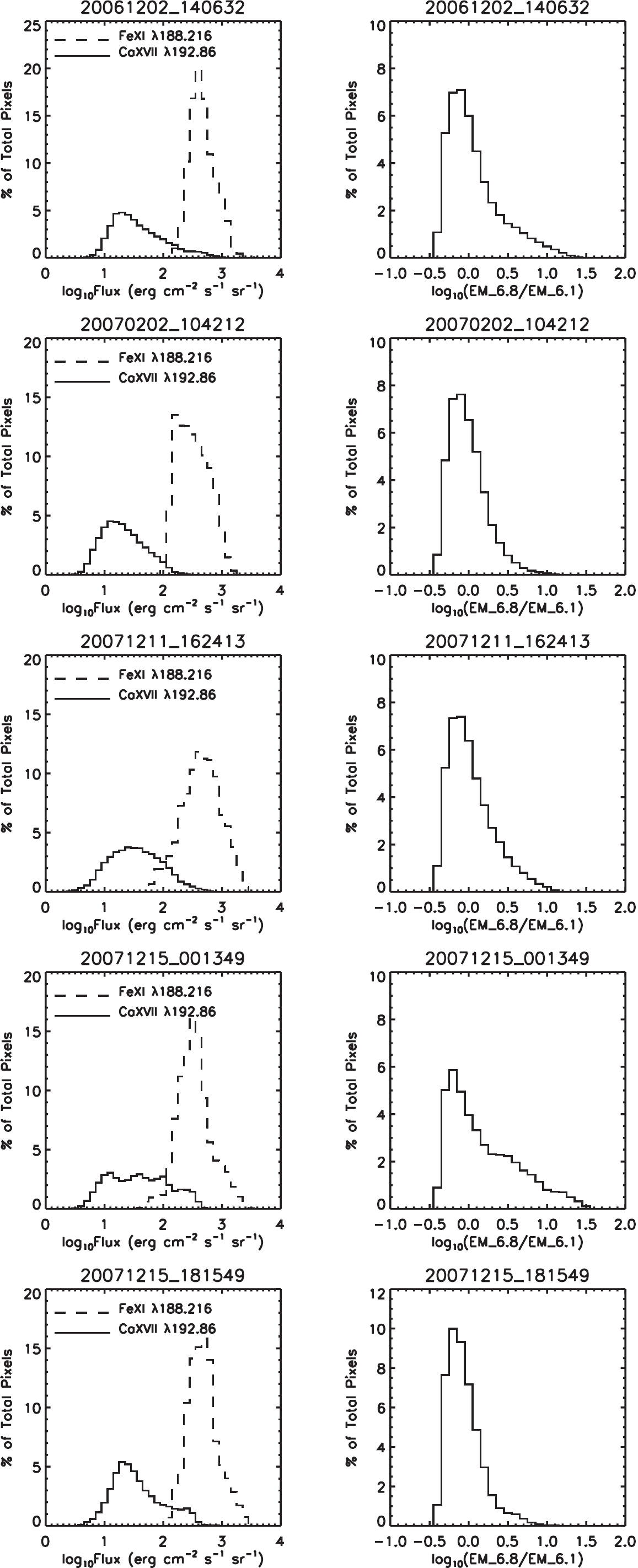}}
\caption{Left panels-- PDF plots for Fe XI $\lambda$188.216 (dashed line) and Ca XVII (solid line) fluxes. Right panels-- PDF plot for the ratio of the emission measure at 6 million degrees to that at 1 million degrees using Eq.(1). For the PDF of Ca XVII fluxes and emission measure ratios, only those pixels with Ca XVII fluxes larger than 10\% of the blend are included.}
\label{fig13}
\end{figure}

\clearpage

\end{document}